\documentclass{aa}
\usepackage{graphicx}
\usepackage{lscape} 
\usepackage{placeins} 
\usepackage{txfonts}
\usepackage{lipsum}
\usepackage{xcolor}
\usepackage{geometry}
\usepackage{hyperref}
\usepackage{subcaption}
\usepackage{soul}
\begin{document}

\title{Joint Observations of PTPS Targets (JOTA). Combined spectroscopic and photometric analysis of 16 SB1 systems}

   \titlerunning{Combined spectroscopic and photometric analysis of 16 SB1 systems.}
   \authorrunning{Jaros, R. et al.}

\author{Jaros R.\inst{1}
\and Niedzielski A. \inst{1}
\and Ali, S.\inst{1}
\and Srivastava D. \inst{1}
\and Sierzputowska J. \inst{1}
\and Adam\'ow, M. \inst{2}
\and Lillo-Box J.\inst{3}
\and Orikhovskyi D. \inst{4}
\and Wolszczan, A.\inst{5}
\and Villaver E. \inst{6, 7}
\and Pribulla T.\inst{4}
\and Va\v{n}ko M.\inst{4}
}

\institute{
    Institute of Astronomy, Nicolaus Copernicus University in Toruń, ul. Gagarina 11, 87-100 Toruń, Poland, \email{jaros@doktorant.umk.pl, Andrzej.Niedzielski@umk.pl}.
\and 
    Center for Astrophysical Surveys, National Center for Supercomputing Applications, 1205 West Clark Street, Urbana, Illinois 61801, USA.
\and 
    Centro de Astrobiología (CAB), CSIC-INTA, Camino Bajo del Castillo s/n, 28692, Villanueva de la Cañada, Madrid, Spain.
\and 
    Astronomical Institute of the Slovak Academy of Sciences, Tatranská Lomnica, SK-059 60 Vysoké Tatry, Slovakia.
\and
    The Pennsylvania State University, Department of Astronomy and Astrophysics, Center for Exoplanets and Habitable Worlds, University Park, PA, United States.
\and
    Instituto de Astrofísica de Canarias, Vía Láctea s/n, 38200 La Laguna, Tenerife, Spain.
\and
    Universidad de La Laguna (ULL), Astrophysics Department, 38206 La Laguna, Tenerife, Spain.
}

\date{}

\abstract
{The Pennsylvania-Toru\'n Planet Search, which operates on a sample of $\sim$1000 northern-hemisphere stars, has been underway since 2004. Stars with radial velocity (RV) amplitudes exceeding 2 km s$^{-1}$  represent a backup subprogram sample dedicated to monitoring binary stars with multiple instruments.}
   {We used almost 21 years of combined RV measurements to search for Doppler signals consistent with stellar or brown dwarf companions and to produce a catalog of both known and previously unpublished binary stars in our planet-search sample. } 
   {We analyzed the combined RV measurements, searching for stellar companions and obtaining orbital solutions for both known and new binary systems. We also searched for periods in available long-term photometric monitoring data from All Sky Automated Survey (ASAS), and applied \textit{Gaia} astrometric data whenever available. }
   {We report the  results of long-term RV monitoring of 16 systems: new detections of low-mass companions to 11 stars and 
 updated orbital elements based on combined sets of our new and literature data for five systems. 
For two objects (TYC 3318-00789-1 and TYC 3318-01538-1), we find evidence of activity or the influence of another unseen companion, possibly due to line-profile variations or unresolved spectral contamination. 
We present true masses for three systems in our sample: TYC 1931-1040-1, TYC 3451-
1449-1, and TYC 3667-1636-1. We obtained these masses using \textit{Gaia} DR3 non-single star data.
}

{}

\maketitle
\nolinenumbers 
\section{Introduction}
The first object identified as a double star was the pair $\nu^1$ and $\nu^2$ Sagittarii, which Ptolemy discovered and described as such around 130 AD.
A better-known pair, Mizar and Alcor, is a visual double star located in the constellation Ursa Major. These stars were described in various cultures around the world and were often used as a test of visual acuity. The Persian astronomer Zakariya al-Qazwini noted the pair in the 13th century. The pair also appears in Japanese mythology, where Alcor was referred to as a ``lifetime star.'' According to this belief, failure to see both stars indicated old age and impending death within one year. 
Later in the 17th century, following the development of the telescope, Benedetto Castelli discovered that Mizar is composed of two elements, A and B. This discovery identified the first known binary star and was wrongly attributed to Jean Baptiste Riccioli in 1650. 
The term ``binary'' was first used by William Herschel in 1802, in his paper ``On the Construction of the Universe,'' to describe two gravitationally bound stars. For most of history, almost all binary stars were studied individually; this changed toward the end of the second millennium with the development of many new instruments. Currently, we know that about half of solar-like stars are in binary or multiple systems.
The multiplicity fraction increases with primary-star mass \citep{1976ApJS...30..273A,1991A&A...248..485D,2013ARA&A..51..269D}.

Observations showed that binary stars are already present in the pre-main-sequence phase \citep{1994ARA&A..32..465M}, leading to the conclusion that multiplicity is a ubiquitous result of stellar formation. Leading theories of binary star formation rely on pre-stellar core fragmentation either before or shortly after the free-fall collapse phase. Two fragmentation mechanisms occur, the first causing fragmentation of the protostellar disk due to rotational instabilities in the protostellar core. The second mechanism occurs at the end of the collapse phase, creating a fast-rotating core that is unstable to axisymmetric perturbations. This process causes the core to rebound into a ring, which then fragments into several components \citep{1994MNRAS.271..999B}.
Stars mostly form in clusters rather than individually \citep{2003ARA&A..41...57L,2010MNRAS.409L..54B}. Consequently, understanding single-core formation mechanisms requires a broader context. Large-scale numerical simulations of collapsing turbulent clouds revealed highly dynamical processes of stellar formation, which commonly result in widely separated binary systems with separation exceeding a few astronomical units (a.u.; \citealt{2012MNRAS.427.2597A,2013ApJ...766...97M,2008MNRAS.385.1820P}). These simulations also exhibit fragmentation, suggesting that these processes do not differ significantly from those described above.

Stars are fundamental to astronomy, and binary and multiple stars play a significant role in the field. Binary stars serve as cosmic laboratories for studying a number of different phenomena. Generally, they allow for accurate measurements of radii, masses, and luminosities. Close binaries (separations $<$ 10 a.u.; \citealt{2010ApJS..190....1R}, \cite{2012ApJ...745...19K}, \cite{2014prpl.conf..267R} constrain the physics of binary evolution and mass transfer. In pairs with neutron stars or black holes, they also allow us to study how gravitational waves induce orbital decay. Because wide binaries have weak gravitational interactions, they provide excellent laboratories for testing gravitational theories, such as MOND (Modified Newtonian dynamics), and hypotheses concerning dark matter \citep{2023ApJ...952..128C,Carrera2012,2025arXiv250420825G}.
In the field of planetary search observations, identifying which stars are in binary systems is crucial for correct signal interpretation. Observed transits can be contaminated by a background eclipsing binary, thereby mimicking possible planetary detection. The presence of stellar companions can affect the existence of planets in the system. This depends on the separations of the stellar components, as close binaries will most likely either have no planetary companion due to tidal forces within the system or have one on a long-period orbit, which would be difficult to detect \citep{2025A&A...700A.106T}. However, in wide binaries, the weak interactions between the stars may not significantly affect the planetary system around either star\citep{2025CoSka..55c..21B}.
Binary stars also play a crucial role in distance measurements. At one step of the distance ladder, eclipsing binaries enable relatively precise (5\%) measurements to approximately 3 Mpc \citep{2019Natur.567..200P}. Cepheids, as standard candles, behave differently in binary systems, which is important to consider when making accurate measurements. Type 1a supernovae are the most utilized standard candles\citep{1997Sci...276.1378N, 1993ApJ...413L.105P}. They are theorized to result from a binary system in which a white dwarf exceeds the Chandrasekhar mass by accreting material from its companion.

Binary stars play a crucial role in all aspects of astronomy, making them valuable tools for astrophysical research.
Spectroscopic binary stars are frequently found in radial-velocity (RV) planet-search projects that operate on random samples of stars (see e.g.,\cite{2019A&A...631A.125K}. 
One such project is the Pennsylvania-Toruń Planet Search (PTPS; \citealt{2008ASPC..398...71N}). Within this project, the initial sample of about 1000 stars was observed with the Hobby-Eberly Telescope and its high-resolution spectrograph between 2004 and 2012. The final sample of 885 best-characterized targets is described in detail in \cite{deka-szymankiewicz_penn_2018}. To save telescope time, observations of targets for which an RV amplitude of over 2 km s$^{-1}$ was detected were usually discontinued after several epochs of data had been gathered. These objects were occasionally observed later as backup targets. Further observations of these spectroscopic binary candidates were carried out with other instruments. Of the 885 PTPS targets in the final sample, multi-epoch RV observations were collected for 819 targets (479 giants, 222 subgiants, and 118 dwarfs). In this sample, 121 stars with amplitudes exceeding 2 km s$^{-1}$ were identified as spectral binary candidates (67 giants, 41 subgiants, and 13 dwarfs).

In this paper, we present results for 16 systems, for which we obtained additional RV observations with the Calar Alto Fiber-fed Échelle (CAFE) instrument at the 2.2m telescope of the Calar Alto Observatory,  supplemented by new observations conducted at Skalnate Pleso Observatory (SPO) of the Astronomical Institute of the Slovak Academy of Sciences (SAS)
with the 1.3m Cassegrain-Nasmyth telescope and the Multi-SIte COntinuous Spectroscopy (MUSICOS) clone spectrograph. 

We also used additional observations of two stars obtained with the High Accuracy RV Planet Searcher in the northern hemisphere (HARPS-N) at the Telescopio Nazionale \textit{Galileo}. For three systems, we provide the true masses of their companions. 

In Sect. 2 we describe observations and collected data.
In Sect. 3 we present details of our data analysis.
Section 4 provides a detailed description of the stellar hosts of all objects.
Section 5 describes the photometric data used.
Finally, in Sect. 6, we present the results of the data analysis.

\section{Observations}

The main dataset comprised 327 epochs of RV measurements extracted from spectroscopic observations  
collected with the Hobby-Eberly Telescope (HET; \citealt{1998SPIE.3352...34R}) and its high resolution spectrograph (HRS; \citealt{tull1998high}).
We conducted the observations in a queue-schedule mode system \citep{2007PASP..119..556S} between 22 January 22004 and 28 June 2013. We used the same instrument configuration and observing method as in \cite{2004ApJ...611L.133C}. The HRS was not pressure- or temperature-controlled, so the best calibration method was the use of the I2 gas-cell technique \citep{1992PASP..104..270M,butler1996attaining}. The spectrograph was fed with a 2'' fiber in high-resolution mode at R = 60 000, with a gas cell (I2) inserted into its optical path. A detailed application of this technique at HET/HRS was described in \cite{nowak2012} and \cite{nowak_planetary-mass_2013}. The collected spectra typically had a high signal-to-noise (S/N) ratio of 200, resulting in RV data with a precision of 5-8 m s$^{-1}$

We conducted follow-up  observations at Calar Alto Observatory with the 2.2-meter telescope  equipped with the  (CAFE) instrument
between 2013-01-06 and 2015-11-16, resulting in 88 RV epochs.  The CAFE instrument has a resolution of R = 62 000 and uses halogen and thorium-argon (ThAr) lamps to calibrate the spectra. The data-reduction and RV-measurement procedures are described in \cite{2013A&A...552A..31A}. 
The resulting RV measurements typically have uncertainties of  $\approx$100 m s$^{-1}$.

More recently, we obtained follow-up observations at the SPO of the Astronomical Institute of the SAS. We acquired 65 epochs using a high-dispersion échelle spectrograph, a clone of MUSICOS, with a resolution range R = 25000 to 38500 \citep{2024CoSka..54b..43P}. We reduced the raw SPO spectra using IRAF (Image Reduction and Analysis Facility)  and calibrated them with spectra from the ThAr lamp and the LED diode. The resulting spectra were 1D continuum-normalized but were not flux-calibrated. The resulting RVs were less precise (uncertainties of >100 m/s); however, for the targets presented here, they proved significant in improving signal detection.
We obtained RVs using the cross-correlation function (CCF) implemented in the Python package $PyAstronomy$. We divided the continuum-normalized 1D spectra into chunks with the same widths as the original échelle grating orders. We used the spectral masks available in the $CERES$ software as correlating spectra. We divided these correlating spectra into segments with the same lengths as the observed spectra. Of the 56 orders, we rejected the first and last few because of spectral cut-offs and telluric lines. This procedure yielded 48 RV values, which we averaged; we used the mean standard deviation of these values as the measurement uncertainty.

For three stars, we obtained additional data with the
  HARPS-N \citealt{2012SPIE.8446E..1VC} fed by the 3.58 m Telescopio Nazionale \textit{Galileo} (TNG), located at the Observatorio del Roque de los Muchachos on La Palma (Spain).  We obtained the spectra as backup observations for the Tracking Advanced Planetary Systems (TAPAS) project \citep{2015A&A...573A..36N,2017A&A...606A..38V}. 
 We used the spectograph's simultaneous ThAr calibration mode and the K2 cross-correlation mask for calibration.
 We used the standard user pipeline, which is based on the weighted cross-correlation function method, to reduce the data and determine the high-precision RV measurements and their uncertainties.  

In total, we collected 513 RV measurements epochs for the 16 systems studied here (Table \ref{RV-DATA-ALL}). Table \ref{HET_obs} summarizes the data used here, including the time span of the observations, the observed RV amplitude, the median RV uncertainty, and the number of epochs.

\begin{table*}
\centering
\caption{RV data used in this paper (excerpt).}
\begin{tabular}{l|llll}
Star (TYC) & MJD [d] & RV [m s$^{-1}$] & $\sigma$RV [m s$^{-1}$] & instrument \\
\hline
1931-01040 & 53701.3 &   2566.2 &   15.3 &          HET \\
1931-01040 & 53703.3 &   2617.8 &   15.9 &          HET \\
1931-01040 & 54516.3 &   3104.2 &   19.1 &          HET \\
1931-01040 & 54523.1 &   3109.2 &   14.8 &          HET \\
1931-01040 & 55327.1 &   2266.6 &   16.0 &          HET \\
1931-01040 & 55522.3 &  -2153.5 &   21.4 &          HET \\ 
 &  & ... &  &  \\  
\end{tabular}
\tablefoot{The full table is available at the CDS.}
\label{RV-DATA-ALL}
\end{table*}

\section{Stellar sample}

\begin{figure}
    \centering
    \includegraphics[width=\hsize]{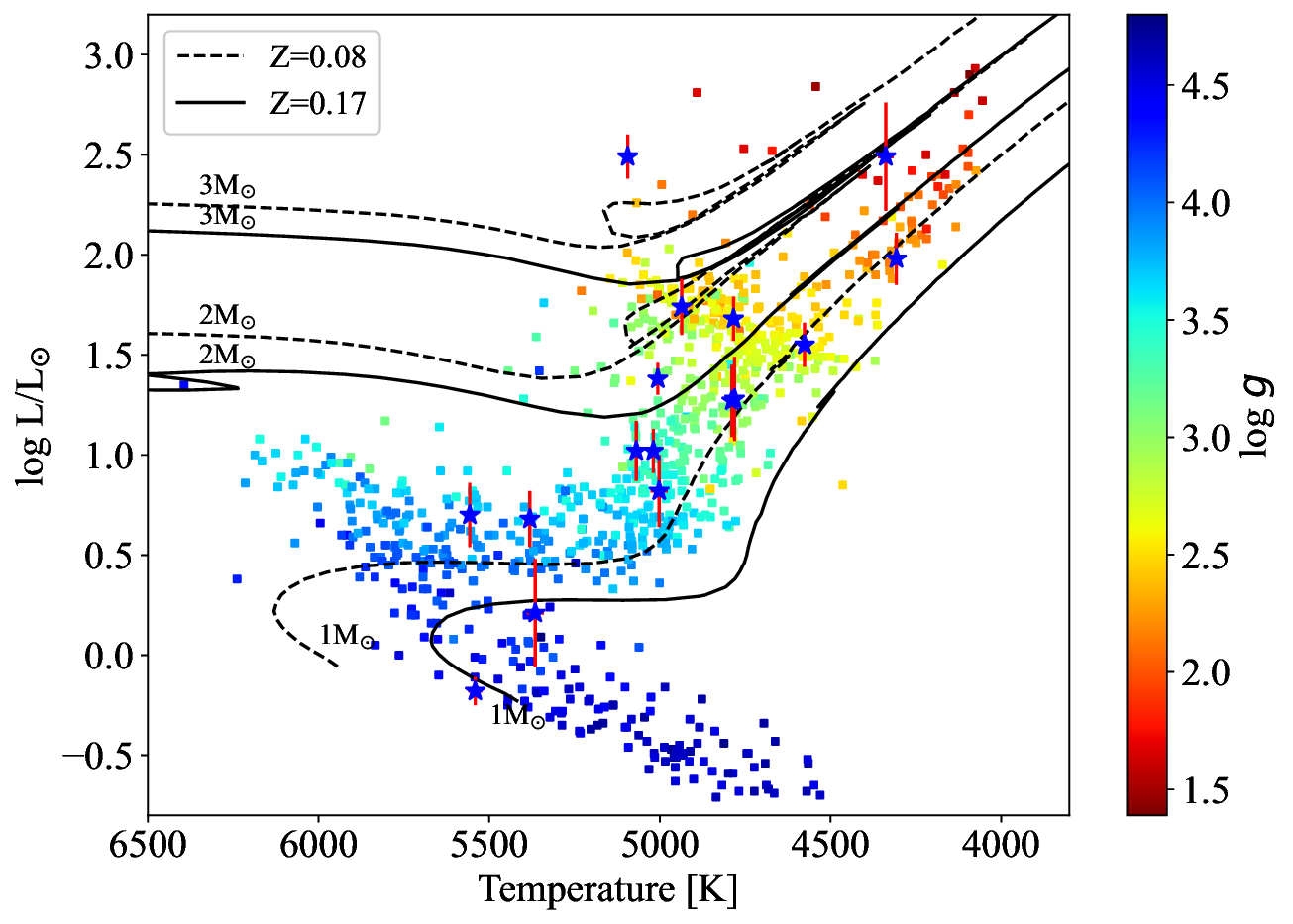}
    \caption{Hertzsprung–Russell diagram  for the complete PTPS sample, with the 16 stars studied here indicated.}
    \label{HRD}
\end{figure}

The sample of stars presented here contains 16 objects from the final PTPS sample \citep{deka-szymankiewicz_penn_2018}, for which atmospheric and integrated stellar parameters were obtained in \cite{2012A&A...547A..91Z}, \cite{2014A&A...569A..55A}, \cite{2016A&A...587A.119A} and summarized in \cite{deka-szymankiewicz_penn_2018}. 
The sample consists of SB1 spectroscopic binaries. We excluded systems with resolved spectral lines (SB2), objects with variable CCFs, and stars with flat CCFs (fast rotators or low-metallicity stars) based on CCF analysis, as described in \cite{2016A&A...585A..73N}.
We list the basic parameters and our estimates of the rotational period in Table 5.
The sample consists of stars for which sufficient observations have been obtained to allow conclusive analyses.  It contains two dwarfs, two subgiants, and 13 giants. Figure \ref{HRD} presents the Hertzsprung-Russell Diagram (HRD)  for the complete PTPS sample, indicating the 16 objects studied here.

\begin{table*}[ht!]
\centering
\caption{Basic data on sample stars \citep{deka-szymankiewicz_penn_2018}.}
\resizebox{\textwidth}{!}{
\begin{tabular}{l c c c c c c c c c c}
\hline
Tycho & Sp type & plx & $T_{eff}$ & log $g$ & [Fe/H] & log $L$  & V$_{rot}sin{i^{\star}}$  & P$_{rot}/sin{i^{\star}}$  & M & R  \\
& & [mas] & [K] & [cm s$^-2$] & [dex] & [$L_{\odot}$] & [km s$^{-1}$] & [d] & [M$_{\odot}$] & [R$_{\odot}$] \\
\hline
0650-01471-1 & G6III & 11.3 $\pm$ 0.9 & 5094 $\pm$ 15 & 2.43 $\pm$ 0.06 & -0.13 $\pm$ 0.02 & 2.49 $\pm$ 0.11 & 3.3 $\pm$ 0.7 & 312 $\pm$ 80 & 3.21 $\pm$ 0.12 & 20.3 $\pm$ 3.7 \\
0697-01743-1 & K0 & 12.0 $\pm$ 1.1 & 5381 $\pm$ 10 & 3.88 $\pm$ 0.04 & 0.24 $\pm$ 0.01 & 0.68 $\pm$ 0.14 & 2.1 $\pm$ 0.6 & 56 $\pm$ 18 & 1.27 $\pm$ 0.06 & 2.33 $\pm$ 0.39 \\
0749-00973-1 & K0 & 28 $\pm$ 12 & 5541 $\pm$ 20 & 4.36 $\pm$ 0.06 & 0.01 $\pm$ 0.03 & -0.18 $\pm$ 0.07 & 0.9 $\pm$ 0.7 & 55 $\pm$ 40 & 0.91 $\pm$ 0.03 & 0.97 $\pm$ 0.16 \\
0870-00084-1 & K0III & 4.0 $\pm$ 2.0 & 5002 $\pm$ 25 & 3.48 $\pm$ 0.09 & -0.29 $\pm$ 0.11 & 0.82 $\pm$ 0.18 & 1 $\pm$ 2 & 210 $\pm$ 520 & 1.06 $\pm$ 0.14 & 3.26 $\pm$ 0.99\\
0870-00937-1 & K0 & 3.0 $\pm$ 1.1 & 4789 $\pm$ 15 & 2.93 $\pm$ 0.06 & -0.04 $\pm$ 0.08 & 1.27 $\pm$ 0.18 & 1.5 $\pm$ 0.7 & 210 $\pm$ 120 & 1.26 $\pm$ 0.22 & 6.3 $\pm$ 1.7\\
1931-01040-1 & G3V & 10.0 $\pm$ 2.1 & 5365 $\pm$ 25 & 4.29 $\pm$ 0.07 & -0.74 $\pm$ 0.03 & 0.21 $\pm$ 0.27 & 3 $\pm$ 2 & 20 $\pm$ 12 & 0.68 $\pm$ 0.02 & 1.22 $\pm$ 0.37 \\
2267-00101-1 & G2V  & 25.5 $\pm$ 3.3 & 5557 $\pm$ 10 & 3.71 $\pm$ 0.03 & -0.50 $\pm$ 0.01 & 0.70 $\pm$ 0.16 & 2.8 $\pm$ 1.3 & 43 $\pm$ 22 & 1.03 $\pm$ 0.06 & 2.38 $\pm$ 0.39 \\
2822-01643-1 & K2 & 3.48 $\pm$ 0.74 & 4307 $\pm$ 15 & 2.36 $\pm$ 0.07 & -0.15 $\pm$ 0.08 & 1.98 $\pm$ 0.13 & 2.5 $\pm$ 0.4 & 295 $\pm$ 84 & 1.10 $\pm$ 0.20 & 14.6 $\pm$ 3.5\\
3018-01050-1 & K1II-III & 3.40 $\pm$ 0.63 & 4576 $\pm$ 18 & 2.69 $\pm$ 0.06 & -0.14 $\pm$ 0.09 & 1.55 $\pm$ 0.11 & 2.3 $\pm$ 0.5 & 191 $\pm$ 59 & 1.11 $\pm$ 0.17 &8.7 $\pm$ 1.9 \\
3314-01371-1 & G5 & 5.0 $\pm$ 1.6 & 5019 $\pm$ 10 & 3.25 $\pm$ 0.04 & -0.06 $\pm$ 0.07 & 1.02 $\pm$ 0.11 & 2.0 $\pm$ 0.8 & 111 $\pm$ 48 & 1.30 $\pm$ 0.17 & 4.39 $\pm$ 0.75 \\
3318-00789-1 & K0 & 3.80 $\pm$ 0.35 & 4781 $\pm$ 13 & 2.82 $\pm$ 0.05 & -0.44 $\pm$ 0.07 & 1.28 $\pm$ 0.21 & 1.9 $\pm$ 0.8 & 167 $\pm$ 75 & 0.94 $\pm$ 0.04 & 6.3 $\pm$ 1.1 \\
3318-01427-1 &  & 2.58 $\pm$ 0.86 & 4784 $\pm$ 10 & 2.64 $\pm$ 0.03 & -0.09 $\pm$ 0.06 & 1.68 $\pm$ 0.11 & 2.0 $\pm$ 0.5 & 252 $\pm$ 74 & 1.54 $\pm$ 0.21 & 10.0 $\pm$ 1.6 \\
3318-01538-1 & K0 & 4.05 $\pm$ 0.91 & 4936 $\pm$ 15 & 2.75 $\pm$ 0.05 & -0.10 $\pm$ 0.09 & 1.74 $\pm$ 0.14 & 2.1 $\pm$ 0.6 & 240 $\pm$ 86 & 1.98 $\pm$ 0.34 & 10.0 $\pm$ 2.2 \\
3319-00172-1 & G6 & 1.42 $\pm$ 0.65 & 4338 $\pm$ 23 & 1.68 $\pm$ 0.09 & -0.45 $\pm$ 0.10 & 2.49 $\pm$ 0.27 & 1.3 $\pm$ 1.1 & 1200 $\pm$ 1200 & 1.57 $\pm$ 0.82 & 31 $\pm$ 14 \\
3451-01449-1 & G8IV & 10.90 $\pm$ 0.39 & 5006 $\pm$ 10 & 3.05 $\pm$ 0.04 & -0.30 $\pm$ 0.01 & 1.38 $\pm$ 0.08 & 0.3 $\pm$ 0.6 & 1300 $\pm$ 3100 & 1.39 $\pm$ 0.04 & 6.18 $\pm$ 0.48 \\
3667-01636-1 & G8III & 6.6 $\pm$ 1.5 & 5069 $\pm$ 15 & 3.25 $\pm$ 0.05 & -0.09 $\pm$ 0.08 & 1.02 $\pm$ 0.15 & 2.0 $\pm$ 0.9 & 109 $\pm$ 55 & 1.32 $\pm$ 0.19 & 4.34 $\pm$ 0.93\\
\hline
\end{tabular}
}

\label{tab:stars}
\end{table*}

\section{Spectroscopic data analysis}

The data analysis process applied here is virtually identical to the procedure we usually follow in our planet-search programs (PTPS and TAPAS).
 This includes analyzing typical stellar-activity indicators  such as the spectral line bisector, the $\mathrm{H}_{\alpha}$ line activity index, or the reversal profile in the cores of Ca H and K lines -- S$_{HK}$, which indicate whether we are dealing with genuine Doppler shifts or spectral line-shape variations caused by stellar-activity-induced phenomena (such as a spot rotating with a star) that mimic such shifts.

\subsection{Periodograms analysis}

Before RV modeling  it is essential to check whether the data contain any genuine signals. We checked the data for outliers, but found none that appeared problematic. We computed Lomb-Scargle (LS; \citealt{Lomb1976}; \citealt{Scargle1982}) and generalized Lomb-Scargle (GLS; \citealt{Zechmeister2009}) periodograms using the Python package $astropy$, which contains both algorithms \citep{2013A&A...558A..33A}.
To assess the statistical significance of the periodogram signals, we plotted false-alarm probability (FAP) levels of 10\%, 1\%, and 0.1\%, on the graphs.

We identified periodic RV signals for most targets. However, for some of the stars studied here, we could not resolve their very long RV variations; therefore, we used estimates only as starting values for subsequent Keplerian modeling.

\subsection{Spectral line bisectors}

We performed an initial check of the origin of the observed RV variations by comparing the RV measurements with the spectral-line bisectors (BIS) available for HET/HRS and TNG/HARPS-N data. The HET/HRS and HARPS-N BIS are calculated from different instruments and sets of
spectral lines and are therefore not directly comparable; we considered them separately.

Variations in the BIS may indicate asymmetries in spectral lines caused by line blending, surface features, oscillations, pulsations, and granulation \citep{2005PASP..117..711G}. Starspots moving with the star's rotation or background binary stars can mimic the RV signal; BIS is a powerful tool for distinguishing these effects from the actual gravitational signal \citep{2001A&A...379..279Q,2002A&A...392..215S}. This method is less sensitive for slowly rotating stars or when the instrumental spectral profiles are significantly broader than the intrinsic profiles \citep{2003A&A...406..373S}. If the signal originates from surface phenomena, such as starspots, an anticorrelation between BIS and RV is expected \citep{2001A&A...379..279Q}. For multiple stars with separations smaller than the spectrograph fiber-width, the expected BIS--RV relations depend on the properties of the components \citep{2015MNRAS.451.2337S,2018MNRAS.478.4720G}. 

Table \ref{BIS} presents the Pearson's correlation coefficients $r$  for RV and BIS, together with the corresponding $p$-values for statistical significance, based on the largest dataset, HET/HRS. For the three stars with available TNG/HARPS-N data, Table \ref{TNG_corr} lists the Pearson's correlation coefficients $r$ for RV-BIS and CCF FWHM, together with the corresponding $p$-values. 

\begin{table}[]

    \caption{Summary of the HET activity analysis.}
    \resizebox{0.45\textwidth}{!}{
    \begin{tabular}{c|cc|cc|cc}
        Tycho & \multicolumn{2}{c|}{BIS} & \multicolumn{2}{c}{$\mathrm{I}_\mathrm{H\alpha}$} & \multicolumn{2}{c}{$\mathrm{I}_\mathrm{Fe}$} \\
        & r & p & r & p & r & p \\
        \hline
        0650-01471-1 & -0.017 & 0.933 & -0.208 & 0.287 & -0.043 & 0.829 \\
        0697-01743-1 & 0.143 & 0.426 & 0.234 & 0.190 & 0.360 & 0.039 \\
        0749-00973-1 & -0.071 & 0.665 & -0.030 & 0.858 & -0.173 & 0.293 \\
        0870-00084-1 & 0.001 & 0.998 & -0.011 & 0.970 & -0.108 & 0.701 \\
        0870-00937-1 & -0.095 & 0.618 & -0.019 & 0.919 & 0.308 & 0.097 \\
        1931-01040-1 & -0.507 & 0.054 & 0.107 & 0.704 & 0.133 & 0.637 \\
        2267-00101-1 & -0.039 & 0.881 & -0.092 & 0.725 & 0.631 & 0.007 \\
        2822-01643-1 & -0.223 & 0.423 & 0.025 & 0.930 & -0.142 & 0.614 \\
        3018-01050-1 & 0.442 & 0.058 & 0.488 & 0.047 & 0.199 & 0.445 \\
        3314-01371-1 & -0.111 & 0.693 & -0.465 & 0.081 & -0.387 & 0.154 \\
        3318-00789-1 & -0.668 & 0.005 & 0.254 & 0.380 & 0.405 & 0.150 \\
        3318-01427-1 & 0.068 & 0.704 & -0.095 & 0.601 & 0.032 & 0.858 \\
        3318-01538-1 & 0.619 & 0.006 & -0.092 & 0.715 & 0.037 & 0.884 \\
        3319-00172-1 & -0.260 & 0.390 & 0.053 & 0.869 & -0.496 & 0.101 \\
        3451-01449-1 & -0.186 & 0.632 & -0.245 & 0.524 & -0.014 & 0.972 \\
        3667-01636-1 & 0.018 & 0.956 & -0.175 & 0.587 & -0.463 & 0.130 \\
        
    \end{tabular}
    }
    \tablefoot{ Pearson's correlation coefficient $r$, with $p$-value of statistical significance for BIS, $\mathrm{I}_\mathrm{H\alpha}$, and $\mathrm{I}_\mathrm{Fe}$. Threshold was set at $p$ = 0.02}
    \label{BIS}
\end{table}

\begin{table*}[]
    \centering
    \caption{A summary of TNG/Harps-N activity analysis.}
    \begin{tabular}{c|cc|cc|cc|cc|cc}
        Tycho & \multicolumn{2}{c|}{BIS} & \multicolumn{2}{c|}{CCF FWHM} & \multicolumn{2}{c|}{$\mathrm{I}_\mathrm{H\alpha}$} & \multicolumn{2}{c|}{$\mathrm{I}_\mathrm{Fe}$} & \multicolumn{2}{c}{$S_\mathrm{HK}$} \\
        & r & p & r & p & r & p & r & p & r & p \\
        \hline
        0749-00973-1 & 0.914 & 0.086 & -0.966 & 0.034 & 0.541 & 0.459 & -0.353 & 0.647 & 0.690 & 0.310 \\
        2267-00101-1 & -0.304 & 0.192 & 0.361 & 0.118 & 0.349 & 0.132 & -0.501 & 0.024 & 0.114 & 0.632 \\
        3314-01371-1 & 0.808 & 0.052 & -0.776 & 0.070 & -0.450 & 0.371 & 0.810 & 0.050 & -0.584 & 0.224 \\        
    \end{tabular}
    \tablefoot{Pearson's correlation coefficient $r$ and the corresponding $p$-values testing the statistical significance for BIS, CCF FWHM, $\mathrm{I}_\mathrm{H\alpha}$, $\mathrm{I}_\mathrm{Fe}$, and $S_\mathrm{HK}$. The threshold was set at $p$ = 0.02}
    \label{TNG_corr}
\end{table*}

\subsection{H$_{\alpha}$ activity index}

In addition to the BIS and activity indicators provided by the TNG/HARPS-N reduction pipeline, we examined the $\mathrm{H}_{\alpha}$ activity index, which is an indicator of chromospheric activity. We measured the $\mathrm{H}_{\alpha}$ activity index in both the HET/HRS and the TNG/HARPS-N spectra using the procedure described in \cite{2013AJ....146..147M} (see also \citealt{2012A&A...541A...9G} or \citealt{2013ApJ...764....3R}).
To control varying instrumental profiles for HET/HRS, we determined an analogical index to $\mathrm{I}_{\mathrm{H\alpha}}$ for Fe I 6593.883~{\AA}. This line is located in the same échelle order as $\mathrm{H}_{\alpha}$ in our HET/HRS data and is unaffected by stellar activity. We denote the Fe activity index by $\mathrm{I}_{\mathrm{Fe}}$. Any correlation between RV and $\mathrm{I}_{\mathrm{Fe}}$ would indicate that the instrumental profile strongly affects the analysis.
Because the two indices are derived from HET/HRS and TNG/HARPS-N spectra with different resolutions, we considered them separately. We present the results of this analysis in Tables \ref{BIS} and \ref{TNG_corr}.

\subsection{Calcium H \& K doublet}

We calculated the reversal profile in the cores of the Ca H and K lines, that is, the emission structure at the cores of the Ca absorption lines, a commonly used indicator of stellar activity \citep{1913ApJ....38..292E}, for every TNG/HARPS-N epoch with $S/N > 10$ in the Ca H \& K region. We refer to this index as S$_{HK}$ in Table \ref{TNG_corr})and calculate it using the formula from \cite{1991ApJS...76..383D}. We calibrated the index against the Mount Wilson scale using the formula provided by \cite{2011arXiv1107.5325L}. The Ca II H \& K lines are not available in HET/HRS spectra.
We present the results in Table \ref{TNG_corr}.

\subsection{Keplerian analysis}

For Keplerian orbit solutions, we analyzed RV data using the Python package RadVel \citep{2018PASP..130d4504F}. After identifying the strongest signal in the generated periodograms based on the FAP levels, we input the initial guesses for the orbital parameters using the available basis ``$P\ T_{c}\ e\ \omega\ K$,'' which is the most standard basis because it is more intuitive for finding the best initial values. However, its disadvantage is that it results in slower Markov chain Monte Carlo (MCMC) processes later in the analysis. The RadVel authors recommend using different bases for fit verification; we used the basis ``$P\ T_{c}\ \sqrt{e}\sin{\omega}\ \sqrt{e}\cos{\omega}\ K$,'' which showed no differences in the fit between bases. For the complete fitting process, we fitted two additional parameters, $\gamma$ and $jitter$. The parameter $\gamma$ represents the instrumental offset required to fit the data with the model. We set $\gamma$ to ``vary'' for all instruments except the HET. We made this choice because the HET RV measurements are relative to an arbitrary point determined by the iodine technique. Consequently, RadVel calculated the HET offset analytically rather than during the subsequent MCMC analysis, and it did not produce an uncertainty for this parameter. 

Here, $jitter$ refers to the so-called white noise, which combines instrumental and astrophysical effects \citep{2018PASP..130d4504F}. We added this term to the RV measurement error and used it as a free parameter to achieve the best $\chi^{2}$.
From the initial fit, we created likelihood objects and then minimized the functions using the $scipy$ package \cite{2020SciPy-NMeth}, applying Nelder-Mead and Powell methods interchangeably depending on the quality of the fit. The ``post'' object formed from this optimization was introduced through the MCMC algorithm to determine the parameter uncertainties.  The MCMC implemented in RadVel is based on the $emcee$ package \citep{2013PASP..125..306F}, which uses the affine-invariant ensemble sampler proposed by \cite{2010CAMCS...5...65G}. To determine whether MCMC found the best minimum value, we used a convergence test. The test is based on the Gelman-Rubin statistics (G-R); when the value of G-R reached unity, we assumed that convergence had occurred. After the burn-in phase, we discarded it and initiated new chains using the burn-in posterior values. We then tested convergence every 50 steps using the methods described in \cite{2013PASP..125...83E}. We halted the MCMC run when the chains were deemed well mixed and the G-R statistic was less than 1.01 for more than 1000 independent samples for each parameter over five consecutive checks. This process results in well-defined posterior values with reasonable uncertainties, which we present in Table \ref{Keplerian}.

\section{Photometry}

We used photometry from the All-Sky Automated Survey for SuperNovae (ASAS-SN), a dedicated, small-aperture network optimized for full-sky, high-cadence transient discovery \citep{Shappee2014,Kochanek2017}. 
Photometry from ASAS-SN provides a long, broadly sampled baseline that is useful for searching for accompanying photometric signals, but several survey characteristics limit its sensitivity to companion-induced effects \cite{Shappee2014,Kochanek2017}. In particular, ASAS-SN light curves typically achieve a photometric precision of a few $10^{-2}$ mag for the brighter stars and degrade toward the survey depth, so the survey is insensitive to very low-amplitude modulations except for the brightest objects. The cadence is effectively daily but irregular, with seasonal visibility gaps. The resulting nontrivial window function produces aliasing that complicates period identification. The cameras have coarse angular resolution and large photometric apertures, which promote blending and flux dilution in crowded fields and thereby reduce sensitivity to shallow transits or other small-amplitude signals. Finally, residual survey systematics such as extinction variations, flat-field residuals, detector artifacts, and other calibration errors can mimic or obscure low-amplitude, long-period trends. Taken together, these limitations imply that only relatively large photometric signatures, e.g. deep eclipses or multi-percent long-term trends, should be robustly detectable in ASAS-SN light curves, while small reflex or reflection signals or shallow transits from long-period companions are generally below the survey’s practical detection threshold.

To assess the presence of periodic variability in our targets, we thoroughly preprocessed each dataset and then computed LS periodograms for all our target ASAS-SN light curves using a customized pipeline. In the following, we detail the step-by-step methodology applied within our photometric pipeline.

\subsection{Proper motion correction}

For several of our targets, the total proper motion is large enough that over the full ASAS-SN observation time span (more than a decade), the projected motion on the sky amounts to a few arcseconds. This motion is a non-negligible fraction of the ASAS-SN pixel scale. In such cases, a fixed aperture centered at the catalog position can produce artificial long-term trends as the stellar image gradually drifts across the detector. To reduce this effect, we followed the strategy recently applied to ASAS-SN photometry by \cite{2025A&A...695A..62S}. This procedure strongly reduced spurious long-term trends associated with the star crossing the detector while preserving the intrinsic variability.

\subsection{Preprocessing and periodicity analyses}

We then filtered these proper-motion-corrected observations to retain the instrumental band with the largest number of measurements. Next, we removed outliers from these filtered measurements using median absolute deviation (MAD) clipping. This effectively removed outliers. We then applied one-day binning to the MAD-clipped data points for each target. 

We then computed a GLS periodogram for this one-day binned time series. We located the peaks with a standard peak-finding routine and ranked them by periodogram power. To mitigate the well-known lunar sampling aliases in ASAS-SN data, we marked the half-monthly and monthly windows $P\in[12,16]$ d and $P\in[27,31]$ d on the periodogram panels and treated peaks within these bands as lunar aliases. The lunar aliases are present in the observations of almost every star in our sample. Lunar-alias peaks are retained in the figures for visual inspection but are excluded when selecting the three highest-ranked astrophysical candidates. If fewer than three non-lunar candidates were available, we used the best remaining peaks and flagged them as lunar. We overlaid the spectral window for each target on each GLS periodogram. We show an example analysis for one of our targets in Fig. \ref{photometry}.As a result of  our analyses, we find no significant photometric periodic variability in the ASAS data for any of our targets.

\section{Results}

We present a combined spectroscopic and photometric variability analysis for a sample of 16  binary systems identified within PTPS, consisting of 11 new binary-star detections (two possibly around active stars), and five independent confirmations of previously reported binary systems.  For the five previously identified systems, we present extended RV data coverage and updated orbital parameters. We also used available literature data for the modeling process. 
We also present the inclinations obtained for three objects from the \textit{Gaia} non-single stars (NSS) catalog \citep{GaiaCollaboration2023NSS}.

Table \ref{Keplerian} presents a summary of the results of the RV analysis described in Sect. 4, where all calculated Keplerian orbital parameters are listed, including the semi-major axis of the orbit in a.u. and the minimum mass in Solar masses.
In Fig. \ref{RV_models} we present the results of Keplerian modeling for the 16 systems, and we briefly describe individual systems below.

\begin{table*}
    \centering
    \caption{Results of Keplerian modeling.}
    \resizebox{\textwidth}{!}{
    \begin{tabular}{c|rrrrrrrr}
    TYC & $P$ [d] & $T_{c}$ [d] & $e$ & $\omega$ [rad]& $K$ [m s$^{-1}$] & $a$ [a.u.] & $m\sin{i}$ [M$_{\odot}$] \\
    \hline
    0697-01743-1 & 3833$\pm$12 & 55777.2$\pm$2.6 & 0.7233$\pm$0.0012 & 5.4715$\pm$0.0024 & 5497$\pm$11 & 5.191$\pm$0.093 & 0.391$\pm$0.012\\
    0749-00973-1 & 4711$\pm$410 & 50294$\pm$650 & 0.569$\pm$0.037 & 4.528$\pm$0.082 & 4346$\pm$550 & 5.33$\pm$0.37 & 0.324$\pm$0.056\\
    0870-00084-1 & 3105$\pm$17 & 51361$\pm$18 & 0.641$\pm$0.019 & 1.393$\pm$ 0.021 & 2179$\pm$64 & 4.2$\pm$0.2 & 0.129$\pm$0.013\\
    0870-00937-1 & 9013$\pm$99 & 53724$\pm$100 & 0.53$\pm$0.02 & 5.347$\pm$0.039 & 2545$\pm$50 & 9.15$\pm$0.59 & 0.282$\pm$0.033\\
    3018-01050-1 & 4565$\pm$24 & 58504$\pm$24 & 0.3451$\pm$0.0022 & 6.363$\pm$0.016 & 4587.6$\pm$8.8 & 5.58$\pm$0.31 & 0.452$\pm$0.043\\
    3314-01371-1 & 844.4$\pm$0.42 & 57067.5$\pm$0.6 & 0.3385$\pm$0.0022 & 1.0304$\pm$0.0064 & 3689$\pm$14 & 1.908$\pm$0.084 & 0.202$\pm$0.018\\
    3318-01427-1 & 5324$\pm$96 & 54296$\pm$2 & 0.5978$\pm$0.0044 & 3.012$\pm$0.004 & 2830$\pm$13 & 6.9$\pm$0.4 & 0.277$\pm$0.026\\
    3319-00172-1 & 4799$\pm$64 & 50220$\pm$76 & 0.463$\pm$0.016 & 4.21$\pm$0.01 & 9549$\pm$49 & 6.5$\pm$1.2 & 1.38$\pm$0.39\\
    3451-01449-1 & 365.941$\pm$0.065 & 56159.6$\pm$2.3 & 0.3381$\pm$0.0058 & 1.892$\pm$0.021 & 6802$\pm$85 & 1.117$\pm$0.011 & 0.3059$\pm$0.0091\\
    \hline
    0650-01471-1 & 1653.49$\pm$0.41 & 51288.6$\pm$1.2 & 0.2966$\pm$0.0014 & 2.5339$\pm$0.0039 & 4417.2$\pm$7.1 & 4.037$\pm$0.051 & 0.568$\pm$0.015 \\
    1931-01040-1 & 811$\pm$1 & 57050.0$\pm$8.1 & 0.172$\pm$0.034 & 3.1$\pm$0.2 & 4003$\pm$110 & 1.497$\pm$0.016 & 0.1529$\pm$0.0066\\
    2822-01643-1 & 8274$\pm$14 & 62466$\pm$17 & 0.2688$\pm$0.0053 & 2.224$\pm$0.015 & 2912$\pm$30 & 8.26$\pm$0.51 & 0.340$\pm$0.042\\
    3667-01636-1 & 1045.96$\pm$0.79 & 57140.5$\pm$3.6 & 0.1496$\pm$0.0027 & 4.81$\pm$0.038 & 3250.4$\pm$6.9 & 2.21$\pm$0.11 & 0.20$\pm$0.02\\

        2267-00101-1 & 700.67$\pm$0.03 & 54951.22$\pm$0.18 & 0.75428$\pm$0.00033 & 5.6854$\pm$0.0014 & 3522.7$\pm$3.7 & 1.56$\pm$0.03 & 0.0983$\pm$0.0039\\
        \hline
        3318-00789-1 & 291.35$\pm$0.045 & 55185$\pm$0.19 & 0.751$\pm$0.002 & 1.855$\pm$0.014 & 9555$\pm$50 & 0.843$\pm$0.012 & 0.2164$\pm$0.0063\\
        3318-01538-1 & 2139$\pm$1 & 58759.4$\pm$2.5 & 0.6774$\pm$0.0021 & 6.1782$\pm$0.0046 & 7100$\pm$51 & 4.08$\pm$0.24 & 0.594$\pm$0.065\\
    \end{tabular}
    }
    \label{Keplerian}
\end{table*}

\subsection{New spectroscopic binaries}

\subsubsection{TYC 0697-1743-1 = HD 32734 = BD+13 797}

TYC 0697-1743-1 is a K0 subgiant with high proper motion\cite{2016ApJS..224...36K},  V=7.88 mag, a slightly supersolar mass of 1.27$\pm$0.06 M$_{\odot}$, and a supersolar metallicity of [Fe/H]=0.24$\pm$0.01 \cite{deka-szymankiewicz_penn_2018}.
Our analysis based on 35 epochs of observations reveals a companion with $m\sin{i}$=0.391$\pm$0.012 M$_{\odot}$ in an orbit with $a$=5.191$\pm$0.093 a.u. ($e$=0.7233$\pm$0.0012).

\subsubsection{TYC 0749-0973-1 = HD 53798 = BD+08 1651}

TYC 0749-0973-1 is a high proper motion star \cite{1915PCinO..18....1P, 2005AJ....129.1483L}, a K0 dwarf with V=8.15, a subsolar mass of 0.91$\pm$0.03 M$_{\odot}$ and a solar metallicity of [Fe/H]=0.01$\pm$0.03 \cite{deka-szymankiewicz_penn_2018}. We gathered 44 epochs of RV data for this star and detected a companion with $m\sin{i}$=0.324$\pm$0.056 M$_{\odot}$ in an orbit with $a$=5.33$\pm$ 0.37 a.u. ($e$=0.569$\pm$0.037).

\subsubsection{TYC 0870-0084-1 = CSI+14-11461 = TIC 14724861}

TYC 0870-0084-1 is a K0 giant with a mass of 1.06$\pm$ 0.14 M$_{\odot}$ and a subsolar metallicity of [Fe/H]=-0.29$\pm$0.11, and a brightness of V=10.22 mag \cite{deka-szymankiewicz_penn_2018} . We collected 21 epochs of RV data for this star and detected a low-mass companion $m\sin{i}$=0.129$\pm$0.013 M$_{\odot}$ in an orbit with $a$=4.2$\pm$0.2 a.u. and $e$=0.641$\pm$0.019.

\subsubsection{TYC 0870-0937-1 = BD+15 2385 = AG+15 1263}

For TYC 0870-0937-1, a giant with a mass of 1.26$\pm$0.22 M$_{\odot}$,  [Fe/H]=-0.04$\pm$0.08 giant, and V=9.82 mag\cite{deka-szymankiewicz_penn_2018}, we collected 35 epochs of RV data. These data indicate the presence of a companion with $m\sin{i}$=0.282$\pm$0.033 M$_{\odot}$ to this star in an extended orbit with $a$=9.15$\pm$0.59 a.u. and $e$=0.53$\pm$0.02.

\subsubsection{TYC 3018-1050-1 = HD 109740 = BD+41 2309}

According to \cite{deka-szymankiewicz_penn_2018} TYC 3018-1050-1 is a giant with a mass of 1.11$\pm$0.17 M$_{\odot}$, a subsolar metallicity of [Fe/H]=-0.14$\pm$0.09, and a brightness of V=8.81 mag. The data presented here, comprising 32 epochs of RV measurements, reveal a companion with $m\sin{i}$=0.452$\pm$0.04 3M$_{\odot}$ in an orbit with $a$=5.58$\pm$0.31 a.u. and $e$=0.3451$\pm$0.0022.

\subsubsection{TYC 3314-1371-1 = Cl Melotte 20 24  =    TIC 116851575}

TYC 3314-1371-1 is a star with V=9.96 mag, a mass of 1.30$\pm$0.17 M$_{\odot}$ and [Fe/H]=-0.06$\pm$0.07 \citep{deka-szymankiewicz_penn_2018} in the $\alpha$ Persei open cluster \cite{1956AN....283..109H}. Our data, comprising 28 epochs of RV measurements, indicate a low-mass companion with $m\sin{i}$=0.202$\pm$0.018 M$_{\odot}$ to this star in an orbit with $a$=1.908$\pm$0.084 a.u. and $e$=0.3385$\pm$0.0022.

\subsubsection{TYC 3318-0789-1  = BD+49 835}

This star was identified as a double or multiple system in \cite{1983BICDS..24...83D}. This is a giant with a mass of 0.94$\pm$0.04 M$_{\odot}$, $\log g$=2.82$\pm$0.05, and 6.3$\pm$1.1 R$_{\odot}$. It has a low metallicity of [Fe/H]=-0.44$\pm$0.07 and a brightness of V=9.52 mag according to \cite{deka-szymankiewicz_penn_2018}. 
We present 22 epochs of RV that show changes in the stellar spectra, which we interpret as Doppler shifts and model as Keplerian motion of the star around the barycenter of the system. These data indicate the presence of an $m\sin{i}$=0.2164$\pm$0.0063 M$_{\odot}$ companion in an eccentric orbit with $a$=0.843$\pm$0.012 a.u. and $e$=0.751$\pm$0.002. 
However, our HET/HRS RVs show a statistically significant anticorrelation between the measured RVs and the spectral-line bisector. Taken alone, this anticorrelation poses a significant problem for identifying the true origin of the observed RV variations and points to either a surface feature rotating with the star or an unresolved companion within the fiber. The apparent orbital period of 291 days is longer than the maximum rotational period (167$\pm$75 d) estimated in \cite{deka-szymankiewicz_penn_2018}. This fact, together with the  observed RV semi-amplitude of more than 9.5 km s$^{-1}$and the apparent lack of photometric variability,
makes a spot hypothesis unlikely. The BIS GLS periodogram in Fig. \ref{rv_activity} shows weak power close to the RV period of $\sim 291$ d, but this power remains below the 10\% FAP level and is not the dominant BIS peak. Therefore, the RV-BIS anticorrelation is unlikely to be caused by stellar activity at the RV period. We therefore conclude that the observed RV-BIS anticorrelation is either spurious, owing to the relatively low number of observations (16 epochs), or caused by a fainter nearby companion to the star that we do not resolve with the HET/HRS fiber. Nevertheless, the binary hypothesis presented here is tentative.

\subsubsection{TYC 3318-1427-1 = TIC 116723654 }

TYC 3318-1427-1, a star with V=9.94 mag, a mass of 1.54$\pm$0.21 M$_{\odot}$, and [Fe/H]=-0.09$\pm$0.06, is a giant with $\log g$=2.64$\pm$0.03 and 10.0$\pm$1.6 R$_{\odot}$ \cite{deka-szymankiewicz_penn_2018}. 
The 39 epochs of RV measurements presented here indicate the presence of a low-mass companion with $m\sin{i}$=0.277$\pm$0.026 M$_{\odot}$ in an extended, eccentric orbit with $a$=6.9$\pm$0.4  a.u. and $e$=0.5978$\pm$0.0044.

\subsubsection{TYC 3318-1538-1  = HD 18927 = BD+49 838}

TYC 3318-1538-1 is a giant with V=8.34 mag, a mass of 1.98$\pm$0.34 M$_{\odot}$, 10.0$\pm$2.2 R$_{\odot}$, $\log g$=2.75$\pm$0.05, and a roughly solar metallicity of [Fh/H]=-0.10$\pm$0.09 \cite{deka-szymankiewicz_penn_2018}.
The 35 epochs of RV presented here, interpreted as Keplerian motion of the star, point to a companion with $m\sin{i}$=0.594$\pm$0.065 in an orbit with $a$=4.08$\pm$0.24 a.u. and $e$=0.6774$\pm$0.0021. However, our HET/HRS data show a statistically significant correlation between RV and BIS. 
 At the same time, we note the lack of a periodic signal in the GLS periodogram for BIS (Fig. \ref{rv_activity}). 
The high amplitude and very long period of the observed RV variations, together with the lack of photometric variability, make the hypothesis of a spot rotating with the star unlikely. Therefore, we either have a spurious correlation owing to the low number of data points (18 in this case) or there is a fainter nearby star that is not resolved within the HET/HRS fiber. The binary hypothesis presented here is therefore tentative.

\subsubsection{TYC 3319-0172-1 = BD+49 872 =  SAO 38637}

TYC 3319-0172-1, another member of the $\alpha$ Persei open cluster in our sample, is a giant star with a mass of 1.57$\pm$0.82 M$_{\odot}$, $\log g$=1.68$\pm$0.09, a low metallicity [Fe/H]=-0.45$\pm$0.10, a large radius of 31 ± 14 R$_{\odot}$, and a brightness of V=9.24 mag \cite{deka-szymankiewicz_penn_2018}. Our data, comprising a modest 17 epochs of RV measurements, indicate the presence of a companion with a similar mass, $m\sin{i}$=1.38$\pm$0.39 M$_{\odot}$, in an orbit with $a$=6.5$\pm$1.2 a.u. and $e$=0.463$\pm$0.016; the companion is not visible in our spectra.

\subsubsection{TYC 3451-1449-1  = HD 100030  = BD+48 1952}

TYC 3451-1449-1 is a high-proper-motion G9IV star \cite{1961csot.book.....L}  \cite{1973AJ.....78...37S}. According to \cite{deka-szymankiewicz_penn_2018} it has V=6.42 mag, a mass of 1.39$\pm$0.04 M$_{\odot}$, [Fe/H]=-0.30$\pm$0.01 and 6.18$\pm$0.48 R$_{\odot}$. Based on the modest 21 epochs of RV measurements collected here, we find an $m\sin{i}$=0.3059$\pm$0.0091 M$_{\odot}$ companion orbiting this star in an orbit with $a$=1.117$\pm$0.011 a.u. and $e$=0.3381$\pm$0.0058.

\subsection{Updates on known systems}

\subsubsection{TYC 0650-1471-1 = o Tau = HD 21120 = BD+08 511}

 \cite{1907ApJ....26..292C} reported RV variations of this star and \cite{1957ApJ...125..712J} subsequently presented the orbital elements for the first time. Additional RV measurements are available in \cite{2008AJ....135..209M}.

According to \cite{deka-szymankiewicz_penn_2018}, TYC 0650-1471-1 is a bright giant with V=3.6 mag, a mass of 3.21$\pm$0.12 M$_{\odot}$, and a subsolar metallicity of [Fe/H]=-0.13$\pm$0.02. It is the most massive star presented here.

Our observations, comprising 32 epochs of RV measurements, combined with data from 
\cite{1957ApJ...125..712J}
and 
\cite{2008AJ....135..209M},
show a companion with $m\sin{i}$=0.568$\pm$0.015 M$_{\odot}$ in an orbit with $a$=4.037$\pm$0.051 a.u. and $e$=0.2966$\pm$0.0014. These results are very similar to those of \cite{1957ApJ...125..712J} but are not identical. Specifically, our eccentricity is slightly larger.

\cite{2019A&A...623A..72K} presents a normalized mass of the o Tau companion based on \textit{Gaia} DR2 proper-motion measurements: 141.49 M$_{J}$AU$^{-1/2}$. 
When applied to our semi-major axis value, this result yields a mass that is too low compared to our msini = 595 M$_{J}$. We note, however, that these authors used a slightly higher mass for the primary star,3.343 M$_\odot$, compared to our value of 3.21 M$_\odot$. Even this difference  is not sufficient to reconcile our minimal value with the literature value.

\subsubsection{TYC 1931-1040-1  = BD+25 1858 = AG+24 940}

\cite{2002AJ....124.1144L}, identified this high-proper-motion star \citep{1955ApJS....2..195R} as a single-lined spectroscopic binary (SB1) and provided both RV data and orbital elements. 
In our analysis we adopted the mass of this main-sequence star, 0.68$\pm$0.02 M$_{\odot}$, as determined in \cite{deka-szymankiewicz_penn_2018}. According to the same study, this is a very low-metallicity star with [Fe/H]=-0.74$\pm$0.03 and a brightness of V=9.65 mag.

Our modeling of  the 30 RV epochs presented here, together with the additional data from \cite{2002AJ....124.1144L}, resulted in the detection of a companion with $m\sin{i}$=0.1529$\pm$0.0066 M$_\odot$ in an orbit with $a$=1.497$\pm$0.016 a.u. and $e$=0.172$\pm$0.034. 
The orbital period derived here is longer than that reported in \cite{2002AJ....124.1144L}, but eccentricities and K agree within the uncertainties.
\cite{2019A&A...623A..72K} present a normalized mass for the companion based on \textit{Gaia} DR2 proper-motion measurements: 254.93 $M_{J}$AU$^{-1/2}$, in agreement with our determination.

\subsubsection{TYC 2267-0101-1  = HD 225239  = BD+33 4828}

This high-velocity star \citep{1955ApJS....2..195R}, initially considered an RV standard, was identified as an RV variable object using the spectrophotometer CORAVEL by \cite{1988A&A...203..329J}. 
It was studied in detail by \cite{2018AJ....156..117F}, who report an orbital period of 700.60$\pm$0.12 days, $K$=3.529$\pm$0.017 km s$^{-1}$, and $e$=0.7582$\pm$0.0020.

Adopting the mass of  
1.03$\pm$0.06 M$_{\odot}$ from \cite{deka-szymankiewicz_penn_2018} for this low-metallicity giant, with[Fe/H]=-0.50$\pm$0.01 and V=6.1 mag, and using our own 48 RV epochs together with all the data available from
\cite{2018AJ....156..117F}, we report a companion with
 $m\sin{i}$=0.0983$\pm$0.0039 M$_{\odot}$ in an orbit with $a$=1.56$\pm$0.03 a.u., P=700.67$\pm$0.03 d,  and $e$=0.75428$\pm$0.00033 orbit.
 This result is consistent with that of \cite{2018AJ....156..117F}.

 However, our spectral analysis reveals an intriguing relation between our HET/HRS and TNG/HARPS-N RV measurements and the shape of the Fe I 6593.883~{\AA} line, which we use to test the stability of the spectral-line profiles. In the HET/HRS data, we find a statistically significant (p$\leq$0.02) RV-$\mathrm{I}_{\mathrm{Fe}}$ correlation, whereas in the case of TNG/HARPS-N data we find a statistically significant anticorrelation. There is no correlation between $\mathrm{I}_{\mathrm{Fe}}$  and $\mathrm{H}_{\alpha}$ in either dataset. 
 There is no periodic signal in the GLS periodogram for $\mathrm{I}_{\mathrm{Fe}}$ (Fig. \ref{rv_activity}).
 This may be a manifestation of another light source unresolved by the 2 arcsec fibers of both instruments. This interpretation is supported to some extent by \cite{2019A&A...631A.125K}.  This potentially giant-BD system merits further study.

\subsubsection{TYC 2822-1643-1 = HD 9519 = BD+41 304}

The orbital solution for the companion TYC 2822-1643 was presented by \cite{2009Obs...129...54G}. This is a giant star with a mass of 1.10$\pm$0.20 M$_{\odot}$, V=7.9 mag, a radius of 
of 14.6$\pm$3.5 R$_{\odot}$, logg=2.36$\pm$ 0.07, and
[Fe/H]=-0.15 ± 0.08 \cite{deka-szymankiewicz_penn_2018}.
We used the original data from\cite{2009Obs...129...54G}, together with our 27 RV measurements, for the Keplerian modeling of this system. We find a companion with
$m\sin{i}$=0.340$\pm$0.042 M$_{\odot}$ orbiting this star with $a$=8.26$\pm$0.51 a.u. and $e$=0.2688$\pm$0.0053. These results are consistent with those of \cite{2009Obs...129...54G}.

\subsubsection{TYC 3667-1636-1  = BD+57 144}

TYC 3667-1636 was identified as a spectroscopic binary by \textit{Gaia} \citep{2019A&A...623A..72K, 2022yCat.1357....0G}.
The orbital solution is present in \textit{Gaia}'s NSS catalog, we discuss it further in Sect. 6.3. 

TYC 3667-1636 is a G8III giant 
 with a mass of 1.32$\pm$0.19 M$_{\odot}$, V=8.86 mag, and
[Fe/H]=-0.09$\pm$0.08 \cite{deka-szymankiewicz_penn_2018}. 
We modeled the 24 RV epochs presented here and detected a companion with $m\sin{i}$ = 0.20$\pm$0.02  M$_{\odot}$ in an orbit with $a$ = 2.21$\pm$0.11 a.u. and $e$ = 0.1496$\pm$0.0027.

\cite{2019A&A...623A..72K} present a normalized mass of 278.92 M$_{J}$AU$^{-1/2}$ for the companion. In this case, the companion's mass exceeds our minimum-mass estimate.

\subsection{True masses}

Three systems in our sample, TYC~1931-1040-1, TYC~3451-1449-1, and TYC~3667-1636-1, are also present in the \textit{Gaia} DR3 NSS catalog. The \textit{Gaia} DR3 was the first \textit{Gaia} release to provide NSS solutions on a large scale, including orbital models listed in the \texttt{nss\_two\_body\_orbit}  table. TYC~1931-1040-1 is also included in the related \textit{Gaia} archive table \texttt{binary\_masses}, which provides component-mass estimates derived from the NSS solutions and adopted assumptions about the stellar components \cite{GaiaCollaboration2023NSS, GaiaCollaboration2023DR3Summary}. We obtained the inclination of the three stars by converting the \textit{Gaia} NSS Thiele-Innes orbital elements provided in the \texttt{nss\_two\_body\_orbit} into the Campbell representation.

For TYC~1931-1040-1, \textit{Gaia} DR3 reports an \texttt{AstroSpectroSB1} solution, combining astrometric and single-lined spectroscopic models, with $P = 737.02 \pm 0.73$~d and $e = 0.141 \pm 0.010$. Our updated RV solution listed in Table~\ref{Keplerian} yields a longer period and slightly higher eccentricity. Using $i = 158.80^\circ \pm 0.69^\circ$, we derive a true companion mass of $0.4229 \pm 0.0225\,M_{\odot}$. For this system, the \textit{Gaia} DR3 \texttt{binary\_masses} table provides an independent companion-mass estimate of $0.380^{+0.050}_{-0.049}\,M_{\odot}$. This value is consistent with ours within the uncertainties.

For TYC~3451-1449-1, our orbital solution given in Table~\ref{Keplerian} agrees closely with the \textit{Gaia} DR3 \texttt{Orbital} solution, which gives $P = 372.72 \pm 0.54$~d and $e = 0.296 \pm 0.022$. Using $i = 73.84^\circ \pm 1.00^\circ$, we derive a true companion mass of $0.3189 \pm 0.0096\,M_{\odot}$.

For TYC~3667-1636-1, \textit{Gaia} provides both \texttt{SB1} and \texttt{Orbital} (astrometric orbital) solutions. The \texttt{SB1} solution gives  $P = 1031.8 \pm 62.1$~d and $e = 0.134 \pm 0.030$, whereas the \texttt{Orbital} solution gives $P = 905.4 \pm 21.1$~d and $e = 0.129 \pm 0.023$. Our orbital solution listed in Table~\ref{Keplerian} is closer to the \texttt{SB1} values. Using the estimated inclination $i = 33.86^\circ \pm 2.82^\circ$, we obtain a true companion mass of $0.3585 \pm 0.0445\,M_{\odot}$.
Those results should be treated with caution because the inclinations provided by \textit{Gaia} are uncertain \citep{2022AJ....164..196W}. Consequently, the true masses we derive are tentative.

\section{Summary and conclusions}
We collected and analyzed 549 epochs of RV data from four different instruments for a sample of 16 stellar objects at various evolutionary stages. 
The sample consists mostly of giants or subgiants, as well as two dwarfs. We present Keplerian modeling of the data as a dynamical interaction between the star and the unseen orbiting companion, using the Python package RadVel to fit Keplerian orbits.  Table \ref{Keplerian} and Fig.
 \ref{RV_models} present the solutions.
We searched for photometric variability in all targets using the ASAS-SN instrument but found no such variability during the observations.

We report the new detection of stellar companions around 11 stars. 
For five known binary stars, we present updated orbital elements based on combined sets of our new data and data from the literature. Our solutions agree within the uncertainties with those in the literature, and we provide improved constraints on the Keplerian parameter uncertainties.
For two objects (TYC 3318-00789-1 and TYC 3318-01538-1), we find evidence of stellar activity or the influence of  another unseen companion, possibly owing to line-profile variations or unresolved spectral contamination. For three systems in our sample, TYC 1931-1040-1, TYC 3451-
1449-1, and TYC 3667-1636-1,  which are included in the \textit{Gaia}
DR3 NSS catalog and have available inclination values, we present their true masses.
Most of the companions presented here are low-mass stars in relatively long-period orbits, except for the companion of TYC 2267-00101-1, whose mass indicates a possible brown dwarf in the brown dwarf desert \cite{2000PASP..112..137M}. The results presented here add nine previously reported nine low-mass companions from the PTPS \cite{2025AcA....75...63N}.

\section*{Data availability}
Table \ref{RV-DATA-ALL} is available only in electronic form through the CDS via anonymous ftp to cdsarc.u-strasbg.fr (130.79.128.5) or via http://cdsweb.u-strasbg.fr/cgi-bin/qcat?J/A+A/

\begin{acknowledgements}
    We want to thank Richard Komzik and Peter Sivanic for observations collected at Skalnate Pleso Observatory. DO, TP, and MV acknowledge support from the Slovak Research and Development Agency (contract
No. APVV-24-0160) and the VEGA grant of the Slovak Academy of Sciences (No. 2/0033/26).
    This research has made use of the SIMBAD database,
operated at CDS, Strasbourg, France. This research has made use of NASA’s Astrophysics Data System. This research made use of Astropy, a community-developed core Python package for Astronomy (Astropy2013). We also want to thank the referee for crucial feedback that improved the clarity of this paper.
\end{acknowledgements}

\bibliographystyle{aa}
\bibliography{bib}

@ARTICLE{1907ApJ....26..292C,
       author = {{Campbell}, W.~W. and {Moore}, J.~H.},
        title = "{Eight stars whose radial velocities vary.}",
      journal = {\apj},
         year = 1907,
        month = nov,
       volume = {26},
        pages = {292-295},
          doi = {10.1086/141506},
       adsurl = {https://ui.adsabs.harvard.edu/abs/1907ApJ....26..292C}
}

@ARTICLE{1913ApJ....38..292E,
       author = {{Eberhard}, G. and {Schwarzschild}, K.},
        title = "{On the reversal of the calcium lines H and K in stellar spectra.}",
      journal = {\apj},
         year = 1913,
        month = oct,
       volume = {38},
        pages = {292-295},
          doi = {10.1086/142037},
       adsurl = {https://ui.adsabs.harvard.edu/abs/1913ApJ....38..292E}
}

@ARTICLE{1915PCinO..18....1P,
       author = {{Porter}, J. and {Yowell}, E. and {Smith}, E.},
        title = "{Part I-IV - Catalogue of Proper Motion Stars}",
      journal = {Publications of the Cincinnati Observatory},
         year = 1915,
        month = jan,
       volume = {18},
        pages = {1-1},
       adsurl = {https://ui.adsabs.harvard.edu/abs/1915PCinO..18....1P}
}

@ARTICLE{1955ApJS....2..195R,
       author = {{Roman}, Nancy G.},
        title = "{A Catalogue of High-Velocity Stars.}",
      journal = {\apjs},
         year = 1955,
        month = dec,
       volume = {2},
        pages = {195},
          doi = {10.1086/190021},
       adsurl = {https://ui.adsabs.harvard.edu/abs/1955ApJS....2..195R}
}

@ARTICLE{1956AN....283..109H,
       author = {{Heckmann}, Otto and {Dieckvoss}, W. and {Kox}, H.},
        title = "{Eigenbewegungen in der Umgebung von {\ensuremath{\alpha}} Persei}",
      journal = {Astronomische Nachrichten},
         year = 1956,
        month = may,
       volume = {283},
        pages = {109},
          doi = {10.1002/asna.19562830211},
       adsurl = {https://ui.adsabs.harvard.edu/abs/1956AN....283..109H}
}

@ARTICLE{1957ApJ...125..712J,
       author = {{Jackson}, E.~S. and {Shane}, W.~W. and {Lynds}, Beverly T.},
        title = "{The Orbits of the Spectroscopic Binaries Omicron Tauri, XI CANCRI, and MU Ursae Majories.}",
      journal = {\apj},
         year = 1957,
        month = may,
       volume = {125},
        pages = {712},
          doi = {10.1086/146345},
       adsurl = {https://ui.adsabs.harvard.edu/abs/1957ApJ...125..712J}
}

@BOOK{1961csot.book.....L,
       author = {{Luyten}, Willem Jacob},
        title = "{A catalogue of 7127 stars in the Northern Hemisphere with proper motions exceeding 0.''2 annually.}",
         year = 1961,
       adsurl = {https://ui.adsabs.harvard.edu/abs/1961csot.book.....L}
}

@ARTICLE{1973AJ.....78...37S,
       author = {{Schild}, R.~E.},
        title = "{Spectral types and UBV photometry of G-K giants at the north galactic pole.}",
      journal = {\aj},
         year = 1973,
        month = feb,
       volume = {78},
        pages = {37-44},
          doi = {10.1086/111369},
       adsurl = {https://ui.adsabs.harvard.edu/abs/1973AJ.....78...37S}
}

@article{Lomb1976,
  author = {Lomb, N. R.},
  title = {Least-squares frequency analysis of unequally spaced data},
  journal = {Astrophysics and Space Science},
  year = {1976},
  volume = {39},
  pages = {447--462}
}

@ARTICLE{1976ApJS...30..273A,
       author = {{Abt}, H.~A. and {Levy}, S.~G.},
        title = "{Multiplicity among solar-type stars.}",
      journal = {\apjs},
         year = 1976,
        month = mar,
       volume = {30},
        pages = {273-306},
          doi = {10.1086/190363},
       adsurl = {https://ui.adsabs.harvard.edu/abs/1976ApJS...30..273A}
}

@article{Scargle1982,
  author = {Scargle, J. D.},
  title = {Studies in astronomical time series analysis. II. Statistical aspects of spectral analysis of unevenly spaced data},
  journal = {The Astrophysical Journal},
  year = {1982},
  volume = {263},
  pages = {835--853}
}

@ARTICLE{1983BICDS..24...83D,
       author = {{Dommanget}, J.},
        title = "{Un catalogue des composantes d'etoiles doubles et multiples (C.C.D.M.)}",
      journal = {Bulletin d'Information du Centre de Donnees Stellaires},
         year = 1983,
        month = mar,
       volume = {24},
        pages = {83},
       adsurl = {https://ui.adsabs.harvard.edu/abs/1983BICDS..24...83D}
}

@ARTICLE{1988A&A...203..329J,
       author = {{Jasniewicz}, G. and {Mayor}, M.},
        title = "{Radial velocity measurements of a sample of northern metal-deficient stars.}",
      journal = {\aap},
         year = 1988,
        month = sep,
       volume = {203},
        pages = {329-340},
       adsurl = {https://ui.adsabs.harvard.edu/abs/1988A&A...203..329J}
}

@ARTICLE{1991A&A...248..485D,
       author = {{Duquennoy}, A. and {Mayor}, M.},
        title = "{Multiplicity among Solar Type Stars in the Solar Neighbourhood - Part Two - Distribution of the Orbital Elements in an Unbiased Sample}",
      journal = {\aap},
         year = 1991,
        month = aug,
       volume = {248},
        pages = {485},
       adsurl = {https://ui.adsabs.harvard.edu/abs/1991A&A...248..485D}
}

@ARTICLE{1991ApJS...76..383D,
       author = {{Duncan}, Douglas K. and {Vaughan}, Arthur H. and {Wilson}, Olin C. and {Preston}, George W. and {Frazer}, James and {Lanning}, Howard and {Misch}, Anthony and {Mueller}, Jean and {Soyumer}, David and {Woodard}, L. and {Baliunas}, Sallie L. and {Noyes}, Robert W. and {Hartmann}, Lee W. and {Porter}, Alain and {Zwaan}, Cornelis and {Middelkoop}, Frans and {Rutten}, Rene G.~M. and {Mihalas}, Dimitri},
        title = "{CA II H and K Measurements Made at Mount Wilson Observatory, 1966--1983}",
      journal = {\apjs},
         year = 1991,
        month = may,
       volume = {76},
        pages = {383},
          doi = {10.1086/191572},
       adsurl = {https://ui.adsabs.harvard.edu/abs/1991ApJS...76..383D}
}

@ARTICLE{1992PASP..104..270M,
       author = {{Marcy}, Geoffrey W. and {Butler}, R.~P.},
        title = "{Precision Radial Velocities with an Iodine Absorption cell}",
      journal = {\pasp},
         year = 1992,
        month = apr,
       volume = {104},
        pages = {270},
          doi = {10.1086/132989},
       adsurl = {https://ui.adsabs.harvard.edu/abs/1992PASP..104..270M}
}

@ARTICLE{1993ApJ...413L.105P,
       author = {{Phillips}, M.~M.},
        title = "{The Absolute Magnitudes of Type IA Supernovae}",
      journal = {\apjl},
         year = 1993,
        month = aug,
       volume = {413},
        pages = {L105},
          doi = {10.1086/186970},
       adsurl = {https://ui.adsabs.harvard.edu/abs/1993ApJ...413L.105P}
}

@ARTICLE{1994MNRAS.271..999B,
       author = {{Bonnell}, I.~A. and {Bate}, M.~R.},
        title = "{The Formation of Close Binary Systems}",
      journal = {\mnras},
         year = 1994,
        month = dec,
       volume = {271},
        pages = {999-1004},
          doi = {10.1093/mnras/271.4.999},
archivePrefix = {arXiv},
       eprint = {astro-ph/9411081},
 primaryClass = {astro-ph},
       adsurl = {https://ui.adsabs.harvard.edu/abs/1994MNRAS.271..999B}
}

@ARTICLE{1994ARA&A..32..465M,
       author = {{Mathieu}, Robert D.},
        title = "{Pre-Main-Sequence Binary Stars}",
      journal = {\araa},
         year = 1994,
        month = jan,
       volume = {32},
        pages = {465-530},
          doi = {10.1146/annurev.aa.32.090194.002341},
       adsurl = {https://ui.adsabs.harvard.edu/abs/1994ARA&A..32..465M}
}

@article{butler1996attaining,
  title={Attaining Doppler precision of 3 M s-1},
  author={Butler, R Paul and Marcy, Geoffrey W and Williams, Eric and McCarthy, Chris and Dosanjh, Preet and Vogt, Steven S},
  journal={Publications of the Astronomical Society of the Pacific},
  volume={108},
  number={724},
  pages={500--500},
  year={1996},
  publisher={The Astronomical Society of the Pacific}
}

@ARTICLE{1997Sci...276.1378N,
       author = {{Nomoto}, K. and {Iwamoto}, K. and {Kishimoto}, N.},
        title = "{Type Ia supernovae: their origin and possible applications in cosmology.}",
      journal = {Science},
         year = 1997,
        month = jan,
       volume = {276},
        pages = {1378-1382},
          doi = {10.1126/science.276.5317.1378},
archivePrefix = {arXiv},
       eprint = {astro-ph/9706007},
 primaryClass = {astro-ph},
       adsurl = {https://ui.adsabs.harvard.edu/abs/1997Sci...276.1378N}
}

@INPROCEEDINGS{1998SPIE.3352...34R,
       author = {{Ramsey}, Lawrence W. and {Adams}, M.~T. and {Barnes}, Thomas G. and {Booth}, John A. and {Cornell}, Mark E. and {Fowler}, James R. and {Gaffney}, Niall I. and {Glaspey}, John W. and {Good}, John M. and {Hill}, Gary J. and {Kelton}, Philip W. and {Krabbendam}, Victor L. and {Long}, L. and {MacQueen}, Phillip J. and {Ray}, Frank B. and {Ricklefs}, Randall L. and {Sage}, J. and {Sebring}, Thomas A. and {Spiesman}, W.~J. and {Steiner}, M.},
        title = "{Early performance and present status of the Hobby-Eberly Telescope}",
    booktitle = {Advanced Technology Optical/IR Telescopes VI},
         year = 1998,
       editor = {{Stepp}, Larry M.},
       series = {Society of Photo-Optical Instrumentation Engineers (SPIE) Conference Series},
       volume = {3352},
        month = aug,
        pages = {34-42},
          doi = {10.1117/12.319287},
       adsurl = {https://ui.adsabs.harvard.edu/abs/1998SPIE.3352...34R}
}

@inproceedings{tull1998high,
  title={High-resolution fiber-coupled spectrograph of the Hobby-Eberly Telescope},
  author={Tull, Robert G},
  booktitle={Optical Astronomical Instrumentation},
  volume={3355},
  pages={387--398},
  year={1998},
  organization={SPIE}
}

@ARTICLE{2001A&A...379..279Q,
       author = {{Queloz}, D. and {Henry}, G.~W. and {Sivan}, J.~P. and {Baliunas}, S.~L. and {Beuzit}, J.~L. and {Donahue}, R.~A. and {Mayor}, M. and {Naef}, D. and {Perrier}, C. and {Udry}, S.},
        title = "{No planet for HD 166435}",
      journal = {\aap},
         year = 2001,
        month = nov,
       volume = {379},
        pages = {279-287},
          doi = {10.1051/0004-6361:20011308},
archivePrefix = {arXiv},
       eprint = {astro-ph/0109491},
 primaryClass = {astro-ph},
       adsurl = {https://ui.adsabs.harvard.edu/abs/2001A&A...379..279Q}
}

@ARTICLE{2002AJ....124.1144L,
       author = {{Latham}, David W. and {Stefanik}, Robert P. and {Torres}, Guillermo and {Davis}, Robert J. and {Mazeh}, Tsevi and {Carney}, Bruce W. and {Laird}, John B. and {Morse}, Jon A.},
        title = "{A Survey of Proper-Motion Stars. XVI. Orbital Solutions for 171 Single-lined Spectroscopic Binaries}",
      journal = {\aj},
         year = 2002,
        month = aug,
       volume = {124},
       number = {2},
        pages = {1144-1161},
          doi = {10.1086/341384},
       adsurl = {https://ui.adsabs.harvard.edu/abs/2002AJ....124.1144L}
}

@ARTICLE{2002A&A...392..215S,
       author = {{Santos}, N.~C. and {Mayor}, M. and {Naef}, D. and {Pepe}, F. and {Queloz}, D. and {Udry}, S. and {Burnet}, M. and {Clausen}, J.~V. and {Helt}, B.~E. and {Olsen}, E.~H. and {Pritchard}, J.~D.},
        title = "{The CORALIE survey for southern extra-solar planets. IX. A 1.3-day period brown dwarf disguised as a planet}",
      journal = {\aap},
         year = 2002,
        month = sep,
       volume = {392},
        pages = {215-229},
          doi = {10.1051/0004-6361:20020876},
archivePrefix = {arXiv},
       eprint = {astro-ph/0206213},
 primaryClass = {astro-ph},
       adsurl = {https://ui.adsabs.harvard.edu/abs/2002A&A...392..215S}
}

@ARTICLE{2003ARA&A..41...57L,
       author = {{Lada}, Charles J. and {Lada}, Elizabeth A.},
        title = "{Embedded Clusters in Molecular Clouds}",
      journal = {\araa},
         year = 2003,
        month = jan,
       volume = {41},
        pages = {57-115},
          doi = {10.1146/annurev.astro.41.011802.094844},
archivePrefix = {arXiv},
       eprint = {astro-ph/0301540},
 primaryClass = {astro-ph},
       adsurl = {https://ui.adsabs.harvard.edu/abs/2003ARA&A..41...57L}
}

@ARTICLE{2003A&A...406..373S,
       author = {{Santos}, N.~C. and {Udry}, S. and {Mayor}, M. and {Naef}, D. and {Pepe}, F. and {Queloz}, D. and {Burki}, G. and {Cramer}, N. and {Nicolet}, B.},
        title = "{The CORALIE survey for southern extra-solar planets. XI. The return of the giant planet orbiting HD 192263}",
      journal = {\aap},
         year = 2003,
        month = jul,
       volume = {406},
        pages = {373-381},
          doi = {10.1051/0004-6361:20030776},
archivePrefix = {arXiv},
       eprint = {astro-ph/0305434},
 primaryClass = {astro-ph},
       adsurl = {https://ui.adsabs.harvard.edu/abs/2003A&A...406..373S}
}

@ARTICLE{2004ApJ...611L.133C,
       author = {{Cochran}, William D. and {Endl}, Michael and {McArthur}, Barbara and {Paulson}, Diane B. and {Smith}, Verne V. and {MacQueen}, Phillip J. and {Tull}, Robert G. and {Good}, John and {Booth}, John and {Shetrone}, Matthew and {Roman}, Brian and {Odewahn}, Stephen and {Deglman}, Frank and {Graver}, Michelle and {Soukup}, Michael and {Villarreal}, Jr., Martin L.},
        title = "{The First Hobby-Eberly Telescope Planet: A Companion to HD 37605}",
      journal = {\apjl},
         year = 2004,
        month = aug,
       volume = {611},
       number = {2},
        pages = {L133-L136},
          doi = {10.1086/423936},
archivePrefix = {arXiv},
       eprint = {astro-ph/0407146},
 primaryClass = {astro-ph},
       adsurl = {https://ui.adsabs.harvard.edu/abs/2004ApJ...611L.133C}
}

@ARTICLE{2005PASP..117..711G,
       author = {{Gray}, David F.},
        title = "{Shapes of Spectral Line Bisectors for Cool Stars}",
      journal = {\pasp},
         year = 2005,
        month = jul,
       volume = {117},
       number = {833},
        pages = {711-720},
          doi = {10.1086/430412},
       adsurl = {https://ui.adsabs.harvard.edu/abs/2005PASP..117..711G}
}

@ARTICLE{2005AJ....129.1483L,
       author = {{L{\'e}pine}, S{\'e}bastien and {Shara}, Michael M.},
        title = "{A Catalog of Northern Stars with Annual Proper Motions Larger than 0.15'' (LSPM-NORTH Catalog)}",
      journal = {\aj},
         year = 2005,
        month = mar,
       volume = {129},
       number = {3},
        pages = {1483-1522},
          doi = {10.1086/427854},
archivePrefix = {arXiv},
       eprint = {astro-ph/0412070},
 primaryClass = {astro-ph},
       adsurl = {https://ui.adsabs.harvard.edu/abs/2005AJ....129.1483L}
}

@ARTICLE{2007PASP..119..556S,
       author = {{Shetrone}, Matthew and {Cornell}, Mark E. and {Fowler}, James R. and {Gaffney}, Niall and {Laws}, Benjamin and {Mader}, Jeff and {Mason}, Cloud and {Odewahn}, Stephen and {Roman}, Brian and {Rostopchin}, Sergey and {Schneider}, Donald P. and {Umbarger}, James and {Westfall}, Amy},
        title = "{Ten Year Review of Queue Scheduling of the Hobby-Eberly Telescope}",
      journal = {\pasp},
         year = 2007,
        month = may,
       volume = {119},
       number = {855},
        pages = {556-566},
          doi = {10.1086/519291},
archivePrefix = {arXiv},
       eprint = {0705.3889},
 primaryClass = {astro-ph},
       adsurl = {https://ui.adsabs.harvard.edu/abs/2007PASP..119..556S}
}

@ARTICLE{2008MNRAS.385.1820P,
       author = {{Price}, Daniel J. and {Bate}, Matthew R.},
        title = "{The effect of magnetic fields on star cluster formation}",
      journal = {\mnras},
         year = 2008,
        month = apr,
       volume = {385},
       number = {4},
        pages = {1820-1834},
          doi = {10.1111/j.1365-2966.2008.12976.x},
archivePrefix = {arXiv},
       eprint = {0801.3293},
 primaryClass = {astro-ph},
       adsurl = {https://ui.adsabs.harvard.edu/abs/2008MNRAS.385.1820P}
}

@INPROCEEDINGS{2008ASPC..398...71N,
       author = {{Niedzielski}, A. and {Wolszczan}, A.},
        title = "{The PennState/Toru{\'n} Center for Astronomy Search for Planets around Evolved Stars.}",
    booktitle = {Extreme Solar Systems},
         year = 2008,
       editor = {{Fischer}, D. and {Rasio}, F.~A. and {Thorsett}, S.~E. and {Wolszczan}, A.},
       series = {Astronomical Society of the Pacific Conference Series},
       volume = {398},
        month = jan,
        pages = {71},
          doi = {10.48550/arXiv.0712.2284},
archivePrefix = {arXiv},
       eprint = {0712.2284},
 primaryClass = {astro-ph},
       adsurl = {https://ui.adsabs.harvard.edu/abs/2008ASPC..398...71N}
}

@ARTICLE{2008AJ....135..209M,
       author = {{Massarotti}, Alessandro and {Latham}, David W. and {Stefanik}, Robert P. and {Fogel}, Jeffrey},
        title = "{Rotational and Radial Velocities for a Sample of 761 HIPPARCOS Giants and the Role of Binarity}",
      journal = {\aj},
         year = 2008,
        month = jan,
       volume = {135},
       number = {1},
        pages = {209-231},
          doi = {10.1088/0004-6256/135/1/209},
       adsurl = {https://ui.adsabs.harvard.edu/abs/2008AJ....135..209M}
}

@ARTICLE{2009Obs...129...54G,
       author = {{Griffin}, R.~F.},
        title = "{Spectroscopic binary orbits from photoelectric radial velocities. Paper 205: HD 9519, delta Aurigae, HR 4427, and HR 7795}",
      journal = {The Observatory},
         year = 2009,
        month = apr,
       volume = {129},
        pages = {54-79},
       adsurl = {https://ui.adsabs.harvard.edu/abs/2009Obs...129...54G}
}

@article{Zechmeister2009,
  author = {Zechmeister, M. and K\"urster, M.},
  title = {The generalised Lomb--Scargle periodogram. A new formalism for the floating-mean and its application to radial-velocity data},
  journal = {Astronomy \& Astrophysics},
  year = {2009},
  volume = {496},
  pages = {577--584}
}

@ARTICLE{2010MNRAS.409L..54B,
       author = {{Bressert}, E. and {Bastian}, N. and {Gutermuth}, R. and {Megeath}, S.~T. and {Allen}, L. and {Evans}, II, Neal J. and {Rebull}, L.~M. and {Hatchell}, J. and {Johnstone}, D. and {Bourke}, T.~L. and {Cieza}, L.~A. and {Harvey}, P.~M. and {Merin}, B. and {Ray}, T.~P. and {Tothill}, N.~F.~H.},
        title = "{The spatial distribution of star formation in the solar neighbourhood: do all stars form in dense clusters?}",
      journal = {\mnras},
         year = 2010,
        month = nov,
       volume = {409},
       number = {1},
        pages = {L54-L58},
          doi = {10.1111/j.1745-3933.2010.00946.x},
archivePrefix = {arXiv},
       eprint = {1009.1150},
 primaryClass = {astro-ph.SR},
       adsurl = {https://ui.adsabs.harvard.edu/abs/2010MNRAS.409L..54B}
}

@ARTICLE{2010CAMCS...5...65G,
       author = {{Goodman}, Jonathan and {Weare}, Jonathan},
        title = "{Ensemble samplers with affine invariance}",
      journal = {Communications in Applied Mathematics and Computational Science},
         year = 2010,
        month = jan,
       volume = {5},
       number = {1},
        pages = {65-80},
          doi = {10.2140/camcos.2010.5.65},
       adsurl = {https://ui.adsabs.harvard.edu/abs/2010CAMCS...5...65G}
}

@ARTICLE{2010ApJS..190....1R,
       author = {{Raghavan}, Deepak and {McAlister}, Harold A. and {Henry}, Todd J. and {Latham}, David W. and {Marcy}, Geoffrey W. and {Mason}, Brian D. and {Gies}, Douglas R. and {White}, Russel J. and {ten Brummelaar}, Theo A.},
        title = "{A Survey of Stellar Families: Multiplicity of Solar-type Stars}",
      journal = {\apjs},
         year = 2010,
        month = sep,
       volume = {190},
       number = {1},
        pages = {1-42},
          doi = {10.1088/0067-0049/190/1/1},
archivePrefix = {arXiv},
       eprint = {1007.0414},
 primaryClass = {astro-ph.SR},
       adsurl = {https://ui.adsabs.harvard.edu/abs/2010ApJS..190....1R}
}

@ARTICLE{2011arXiv1107.5325L,
       author = {{Lovis}, C. and {Dumusque}, X. and {Santos}, N.~C. and {Bouchy}, F. and {Mayor}, M. and {Pepe}, F. and {Queloz}, D. and {S{\'e}gransan}, D. and {Udry}, S.},
        title = "{The HARPS search for southern extra-solar planets. XXXI. Magnetic activity cycles in solar-type stars: statistics and impact on precise radial velocities}",
      journal = {arXiv e-prints},
         year = 2011,
        month = jul,
          eid = {arXiv:1107.5325},
        pages = {arXiv:1107.5325},
          doi = {10.48550/arXiv.1107.5325},
archivePrefix = {arXiv},
       eprint = {1107.5325},
 primaryClass = {astro-ph.SR},
       adsurl = {https://ui.adsabs.harvard.edu/abs/2011arXiv1107.5325L}
}

@phdthesis{nowak2012,
  author       = {Nowak, Grzegorz},
  title        = {PhD Thesis},
  school       = {Nicolaus Copernicus University},
  address      = {Torun, Poland},
  year         = {2012}
}

@ARTICLE{2012MNRAS.427.2597A,
       author = {{Ayliffe}, Ben A. and {Bate}, Matthew R.},
        title = "{The growth and hydrodynamic collapse of a protoplanet envelope}",
      journal = {\mnras},
         year = 2012,
        month = dec,
       volume = {427},
       number = {3},
        pages = {2597-2612},
          doi = {10.1111/j.1365-2966.2012.21979.x},
archivePrefix = {arXiv},
       eprint = {1208.5513},
 primaryClass = {astro-ph.EP},
       adsurl = {https://ui.adsabs.harvard.edu/abs/2012MNRAS.427.2597A}
}

@article{Carrera2012,
    author = {{Carrera}, Daniel},
    title = "{The effect of dark matter capture
on binary stars}",
    journal = {Lund Observatory, Masters Thesis},
    year = 2012
}

@INPROCEEDINGS{2012SPIE.8446E..1VC,
       author = {{Cosentino}, Rosario and {Lovis}, Christophe and {Pepe}, Francesco and {Collier Cameron}, Andrew and {Latham}, David W. and {Molinari}, Emilio and {Udry}, Stephane and {Bezawada}, Naidu and {Black}, Martin and {Born}, Andy and {Buchschacher}, Nicolas and {Charbonneau}, Dave and {Figueira}, Pedro and {Fleury}, Michel and {Galli}, Alberto and {Gallie}, Angus and {Gao}, Xiaofeng and {Ghedina}, Adriano and {Gonzalez}, Carlos and {Gonzalez}, Manuel and {Guerra}, Jose and {Henry}, David and {Horne}, Keith and {Hughes}, Ian and {Kelly}, Dennis and {Lodi}, Marcello and {Lunney}, David and {Maire}, Charles and {Mayor}, Michel and {Micela}, Giusi and {Ordway}, Mark P. and {Peacock}, John and {Phillips}, David and {Piotto}, Giampaolo and {Pollacco}, Don and {Queloz}, Didier and {Rice}, Ken and {Riverol}, Carlos and {Riverol}, Luis and {San Juan}, Jose and {Sasselov}, Dimitar and {Segransan}, Damien and {Sozzetti}, Alessandro and {Sosnowska}, Danuta and {Stobie}, Brian and {Szentgyorgyi}, Andrew and {Vick}, Andy and {Weber}, Luc},
        title = "{Harps-N: the new planet hunter at TNG}",
    booktitle = {Ground-based and Airborne Instrumentation for Astronomy IV},
         year = 2012,
       editor = {{McLean}, Ian S. and {Ramsay}, Suzanne K. and {Takami}, Hideki},
       series = {Society of Photo-Optical Instrumentation Engineers (SPIE) Conference Series},
       volume = {8446},
        month = sep,
          eid = {84461V},
        pages = {84461V},
          doi = {10.1117/12.925738},
       adsurl = {https://ui.adsabs.harvard.edu/abs/2012SPIE.8446E..1VC}
}

@ARTICLE{2012A&A...547A..91Z,
       author = {{Zieli{\'n}ski}, P. and {Niedzielski}, A. and {Wolszczan}, A. and {Adam{\'o}w}, M. and {Nowak}, G.},
        title = "{The Penn State-Toru{\'n} Centre for Astronomy Planet Search stars. I. Spectroscopic analysis of 348 red giants}",
      journal = {\aap},
         year = 2012,
        month = nov,
       volume = {547},
          eid = {A91},
        pages = {A91},
          doi = {10.1051/0004-6361/201117775},
archivePrefix = {arXiv},
       eprint = {1206.6276},
 primaryClass = {astro-ph.EP},
       adsurl = {https://ui.adsabs.harvard.edu/abs/2012A&A...547A..91Z}
}

@ARTICLE{2012A&A...541A...9G,
       author = {{Gomes da Silva}, J. and {Santos}, N.~C. and {Bonfils}, X. and {Delfosse}, X. and {Forveille}, T. and {Udry}, S. and {Dumusque}, X. and {Lovis}, C.},
        title = "{Long-term magnetic activity of a sample of M-dwarf stars from the HARPS program . II. Activity and radial velocity}",
      journal = {\aap},
         year = 2012,
        month = may,
       volume = {541},
          eid = {A9},
        pages = {A9},
          doi = {10.1051/0004-6361/201118598},
archivePrefix = {arXiv},
       eprint = {1202.1564},
 primaryClass = {astro-ph.SR},
       adsurl = {https://ui.adsabs.harvard.edu/abs/2012A&A...541A...9G}
}

@ARTICLE{2012ApJ...745...19K,
       author = {{Kraus}, Adam L. and {Ireland}, Michael J. and {Hillenbrand}, Lynne A. and {Martinache}, Frantz},
        title = "{The Role of Multiplicity in Disk Evolution and Planet Formation}",
      journal = {\apj},
         year = 2012,
        month = jan,
       volume = {745},
       number = {1},
          eid = {19},
        pages = {19},
          doi = {10.1088/0004-637X/745/1/19},
archivePrefix = {arXiv},
       eprint = {1109.4141},
 primaryClass = {astro-ph.EP},
       adsurl = {https://ui.adsabs.harvard.edu/abs/2012ApJ...745...19K}
}

@ARTICLE{2013PASP..125...83E,
       author = {{Eastman}, Jason and {Gaudi}, B. Scott and {Agol}, Eric},
        title = "{EXOFAST: A Fast Exoplanetary Fitting Suite in IDL}",
      journal = {\pasp},
         year = 2013,
        month = jan,
       volume = {125},
       number = {923},
        pages = {83},
          doi = {10.1086/669497},
archivePrefix = {arXiv},
       eprint = {1206.5798},
 primaryClass = {astro-ph.IM},
       adsurl = {https://ui.adsabs.harvard.edu/abs/2013PASP..125...83E}
}

@ARTICLE{2013ApJ...764....3R,
       author = {{Robertson}, Paul and {Endl}, Michael and {Cochran}, William D. and {Dodson-Robinson}, Sarah E.},
        title = "{H{\ensuremath{\alpha}} Activity of Old M Dwarfs: Stellar Cycles and Mean Activity Levels for 93 Low-mass Stars in the Solar Neighborhood}",
      journal = {\apj},
         year = 2013,
        month = feb,
       volume = {764},
       number = {1},
          eid = {3},
        pages = {3},
          doi = {10.1088/0004-637X/764/1/3},
archivePrefix = {arXiv},
       eprint = {1211.6091},
 primaryClass = {astro-ph.SR},
       adsurl = {https://ui.adsabs.harvard.edu/abs/2013ApJ...764....3R}
}

@ARTICLE{2013AJ....146..147M,
       author = {{Maciejewski}, G. and {Niedzielski}, A. and {Wolszczan}, A. and {Nowak}, G. and {Neuh{\"a}user}, R. and {Winn}, J.~N. and {Deka}, B. and {Adam{\'o}w}, M. and {G{\'o}recka}, M. and {Fern{\'a}ndez}, M. and {Aceituno}, F.~J. and {Ohlert}, J. and {Errmann}, R. and {Seeliger}, M. and {Dimitrov}, D. and {Latham}, D.~W. and {Esquerdo}, G.~A. and {McKnight}, L. and {Holman}, M.~J. and {Jensen}, E.~L.~N. and {Kramm}, U. and {Pribulla}, T. and {Raetz}, St. and {Schmidt}, T.~O.~B. and {Ginski}, Ch. and {Mottola}, S. and {Hellmich}, S. and {Adam}, Ch. and {Gilbert}, H. and {Mugrauer}, M. and {Saral}, G. and {Popov}, V. and {Raetz}, M.},
        title = "{Constraints on a Second Planet in the WASP-3 System}",
      journal = {\aj},
         year = 2013,
        month = dec,
       volume = {146},
       number = {6},
          eid = {147},
        pages = {147},
          doi = {10.1088/0004-6256/146/6/147},
archivePrefix = {arXiv},
       eprint = {1309.6733},
 primaryClass = {astro-ph.EP},
       adsurl = {https://ui.adsabs.harvard.edu/abs/2013AJ....146..147M}
}

@misc{nowak_planetary-mass_2013,
	title = {Planetary-mass companions to the {K}-giants {BD}+15 2940 and {HD} 233604},
	url = {http://arxiv.org/abs/1304.6755},
	doi = {10.48550/arXiv.1304.6755},
	urldate = {2025-10-29},
	publisher = {arXiv},
	author = {Nowak, G. and Niedzielski, A. and Wolszczan, A. and Adamów, M. and Maciejewski, G.},
	month = apr,
	year = {2013},
	note = {arXiv:1304.6755},
}

@ARTICLE{2013ARA&A..51..269D,
       author = {{Duch{\^e}ne}, Gaspard and {Kraus}, Adam},
        title = "{Stellar Multiplicity}",
      journal = {\araa},
         year = 2013,
        month = aug,
       volume = {51},
       number = {1},
        pages = {269-310},
          doi = {10.1146/annurev-astro-081710-102602},
archivePrefix = {arXiv},
       eprint = {1303.3028},
 primaryClass = {astro-ph.SR},
       adsurl = {https://ui.adsabs.harvard.edu/abs/2013ARA&A..51..269D}
}

@ARTICLE{2013ApJ...766...97M,
       author = {{Myers}, Andrew T. and {McKee}, Christopher F. and {Cunningham}, Andrew J. and {Klein}, Richard I. and {Krumholz}, Mark R.},
        title = "{The Fragmentation of Magnetized, Massive Star-forming Cores with Radiative Feedback}",
      journal = {\apj},
         year = 2013,
        month = apr,
       volume = {766},
       number = {2},
          eid = {97},
        pages = {97},
          doi = {10.1088/0004-637X/766/2/97},
archivePrefix = {arXiv},
       eprint = {1211.3467},
 primaryClass = {astro-ph.SR},
       adsurl = {https://ui.adsabs.harvard.edu/abs/2013ApJ...766...97M}
}

@ARTICLE{2013A&A...552A..31A,
       author = {{Aceituno}, J. and {S{\'a}nchez}, S.~F. and {Grupp}, F. and {Lillo}, J. and {Hern{\'a}n-Obispo}, M. and {Benitez}, D. and {Montoya}, L.~M. and {Thiele}, U. and {Pedraz}, S. and {Barrado}, D. and {Dreizler}, S. and {Bean}, J.},
        title = "{CAFE: Calar Alto Fiber-fed {\'E}chelle spectrograph}",
      journal = {\aap},
         year = 2013,
        month = apr,
       volume = {552},
          eid = {A31},
        pages = {A31},
          doi = {10.1051/0004-6361/201220361},
archivePrefix = {arXiv},
       eprint = {1301.2066},
 primaryClass = {astro-ph.IM},
       adsurl = {https://ui.adsabs.harvard.edu/abs/2013A&A...552A..31A}
}

@ARTICLE{2013A&A...558A..33A,
       author = {{Astropy Collaboration} and {Robitaille}, Thomas P. and {Tollerud}, Erik J. and {Greenfield}, Perry and {Droettboom}, Michael and {Bray}, Erik and {Aldcroft}, Tom and {Davis}, Matt and {Ginsburg}, Adam and {Price-Whelan}, Adrian M. and {Kerzendorf}, Wolfgang E. and {Conley}, Alexander and {Crighton}, Neil and {Barbary}, Kyle and {Muna}, Demitri and {Ferguson}, Henry and {Grollier}, Fr{\'e}d{\'e}ric and {Parikh}, Madhura M. and {Nair}, Prasanth H. and {Unther}, Hans M. and {Deil}, Christoph and {Woillez}, Julien and {Conseil}, Simon and {Kramer}, Roban and {Turner}, James E.~H. and {Singer}, Leo and {Fox}, Ryan and {Weaver}, Benjamin A. and {Zabalza}, Victor and {Edwards}, Zachary I. and {Azalee Bostroem}, K. and {Burke}, D.~J. and {Casey}, Andrew R. and {Crawford}, Steven M. and {Dencheva}, Nadia and {Ely}, Justin and {Jenness}, Tim and {Labrie}, Kathleen and {Lim}, Pey Lian and {Pierfederici}, Francesco and {Pontzen}, Andrew and {Ptak}, Andy and {Refsdal}, Brian and {Servillat}, Mathieu and {Streicher}, Ole},
        title = "{Astropy: A community Python package for astronomy}",
      journal = {\aap},
         year = 2013,
        month = oct,
       volume = {558},
          eid = {A33},
        pages = {A33},
          doi = {10.1051/0004-6361/201322068},
archivePrefix = {arXiv},
       eprint = {1307.6212},
 primaryClass = {astro-ph.IM},
       adsurl = {https://ui.adsabs.harvard.edu/abs/2013A&A...558A..33A}
}

@ARTICLE{2013PASP..125..306F,
       author = {{Foreman-Mackey}, Daniel and {Hogg}, David W. and {Lang}, Dustin and {Goodman}, Jonathan},
        title = "{emcee: The MCMC Hammer}",
      journal = {\pasp},
         year = 2013,
        month = mar,
       volume = {125},
       number = {925},
        pages = {306},
          doi = {10.1086/670067},
archivePrefix = {arXiv},
       eprint = {1202.3665},
 primaryClass = {astro-ph.IM},
       adsurl = {https://ui.adsabs.harvard.edu/abs/2013PASP..125..306F}
}

@INPROCEEDINGS{2014prpl.conf..267R,
       author = {{Reipurth}, B. and {Clarke}, C.~J. and {Boss}, A.~P. and {Goodwin}, S.~P. and {Rodr{\'\i}guez}, L.~F. and {Stassun}, K.~G. and {Tokovinin}, A. and {Zinnecker}, H.},
        title = "{Multiplicity in Early Stellar Evolution}",
    booktitle = {Protostars and Planets VI},
         year = 2014,
       editor = {{Beuther}, Henrik and {Klessen}, Ralf S. and {Dullemond}, Cornelis P. and {Henning}, Thomas},
        month = jan,
        pages = {267-290},
          doi = {10.2458/azu_uapress_9780816531240-ch012},
archivePrefix = {arXiv},
       eprint = {1403.1907},
 primaryClass = {astro-ph.SR},
       adsurl = {https://ui.adsabs.harvard.edu/abs/2014prpl.conf..267R}
}

@article{Shappee2014,
  author  = {Shappee, B. J. and Prieto, J. L. and Stanek, K. Z. and Holoien, T. W.-S. and
             Thompson, T. A. and Kochanek, C. S. and Dong, S. and Adams, S. M. and et~al.},
  title   = {{The Man Behind the Curtain: X-Rays Drive the UV through NIR Variability
             in the 2013 Active Galactic Nucleus Outburst in NGC 2617}},
  journal = {The Astrophysical Journal},
  volume  = {788},
  number  = {1},
  pages   = {48},
  year    = {2014},
  doi     = {10.1088/0004-637X/788/1/48}
}

@ARTICLE{2014A&A...569A..55A,
       author = {{Adam{\'o}w}, M. and {Niedzielski}, A. and {Villaver}, E. and {Wolszczan}, A. and {Nowak}, G.},
        title = "{The Penn State - Toru{\'n} Centre for Astronomy Planet Search stars. II. Lithium abundance analysis of the red giant clump sample}",
      journal = {\aap},
         year = 2014,
        month = sep,
       volume = {569},
          eid = {A55},
        pages = {A55},
          doi = {10.1051/0004-6361/201423400},
archivePrefix = {arXiv},
       eprint = {1407.4956},
 primaryClass = {astro-ph.SR},
       adsurl = {https://ui.adsabs.harvard.edu/abs/2014A&A...569A..55A}
}

@ARTICLE{2015A&A...573A..36N,
       author = {{Niedzielski}, A. and {Villaver}, E. and {Wolszczan}, A. and {Adam{\'o}w}, M. and {Kowalik}, K. and {Maciejewski}, G. and {Nowak}, G. and {Garc{\'\i}a-Hern{\'a}ndez}, D.~A. and {Deka}, B. and {Adamczyk}, M.},
        title = "{Tracking Advanced Planetary Systems (TAPAS) with HARPS-N . I. A multiple planetary system around the red giant star TYC 1422-614-1}",
      journal = {\aap},
         year = 2015,
        month = jan,
       volume = {573},
          eid = {A36},
        pages = {A36},
          doi = {10.1051/0004-6361/201424399},
archivePrefix = {arXiv},
       eprint = {1410.5971},
 primaryClass = {astro-ph.EP},
       adsurl = {https://ui.adsabs.harvard.edu/abs/2015A&A...573A..36N}
}

@ARTICLE{2015MNRAS.451.2337S,
       author = {{Santerne}, A. and {D{\'\i}az}, R.~F. and {Almenara}, J.-M. and {Bouchy}, F. and {Deleuil}, M. and {Figueira}, P. and {H{\'e}brard}, G. and {Moutou}, C. and {Rodionov}, S. and {Santos}, N.~C.},
        title = "{PASTIS: Bayesian extrasolar planet validation - II. Constraining exoplanet blend scenarios using spectroscopic diagnoses}",
      journal = {\mnras},
         year = 2015,
        month = aug,
       volume = {451},
       number = {3},
        pages = {2337-2351},
          doi = {10.1093/mnras/stv1080},
archivePrefix = {arXiv},
       eprint = {1505.02663},
 primaryClass = {astro-ph.EP},
       adsurl = {https://ui.adsabs.harvard.edu/abs/2015MNRAS.451.2337S}
}

@ARTICLE{2016ApJS..224...36K,
       author = {{Kirkpatrick}, J. Davy and {Kellogg}, Kendra and {Schneider}, Adam C. and {Fajardo-Acosta}, Sergio and {Cushing}, Michael C. and {Greco}, Jennifer and {Mace}, Gregory N. and {Gelino}, Christopher R. and {Wright}, Edward L. and {Eisenhardt}, Peter R.~M. and {Stern}, Daniel and {Faherty}, Jacqueline K. and {Sheppard}, Scott S. and {Lansbury}, George B. and {Logsdon}, Sarah E. and {Martin}, Emily C. and {McLean}, Ian S. and {Schurr}, Steven D. and {Cutri}, Roc M. and {Conrow}, Tim},
        title = "{The AllWISE Motion Survey, Part 2}",
      journal = {\apjs},
         year = 2016,
        month = jun,
       volume = {224},
       number = {2},
          eid = {36},
        pages = {36},
          doi = {10.3847/0067-0049/224/2/36},
archivePrefix = {arXiv},
       eprint = {1603.08040},
 primaryClass = {astro-ph.SR},
       adsurl = {https://ui.adsabs.harvard.edu/abs/2016ApJS..224...36K}
}

@ARTICLE{2016A&A...587A.119A,
       author = {{Adamczyk}, M. and {Deka-Szymankiewicz}, B. and {Niedzielski}, A.},
        title = "{Masses and luminosities for 342 stars from the PennState-Toru{\'n} Centre for Astronomy Planet Search}",
      journal = {\aap},
         year = 2016,
        month = mar,
       volume = {587},
          eid = {A119},
        pages = {A119},
          doi = {10.1051/0004-6361/201526628},
archivePrefix = {arXiv},
       eprint = {1510.07495},
 primaryClass = {astro-ph.SR},
       adsurl = {https://ui.adsabs.harvard.edu/abs/2016A&A...587A.119A}
}

@article{Kochanek2017,
  author  = {Kochanek, C.~S. and Shappee, B.~J. and Stanek, K.~Z. and Holoien, T.~W.-S. and
             Thompson, T.~A. and Prieto, J.~L. and et~al.},
  title   = {{The All-Sky Automated Survey for SuperNovae (ASAS-SN) Light Curve Server v1.0}},
  journal = {Publications of the Astronomical Society of the Pacific},
  volume  = {129},
  number  = {975},
  pages   = {104502},
  year    = {2017},
  doi     = {10.1088/1538-3873/aa80d9}
}

@ARTICLE{2017A&A...606A..38V,
       author = {{Villaver}, E. and {Niedzielski}, A. and {Wolszczan}, A. and {Nowak}, G. and {Kowalik}, K. and {Adam{\'o}w}, M. and {Maciejewski}, G. and {Deka-Szymankiewicz}, B. and {Maldonado}, J.},
        title = "{Tracking Advanced Planetary Systems (TAPAS) with HARPS-N. V. A Massive Jupiter orbiting the very-low-metallicity giant star BD+03 2562 and a possible planet around HD 103485}",
      journal = {\aap},
         year = 2017,
        month = oct,
       volume = {606},
          eid = {A38},
        pages = {A38},
          doi = {10.1051/0004-6361/201730471},
archivePrefix = {arXiv},
       eprint = {1706.01278},
 primaryClass = {astro-ph.EP},
       adsurl = {https://ui.adsabs.harvard.edu/abs/2017A&A...606A..38V}
}

@ARTICLE{2018PASP..130d4504F,
       author = {{Fulton}, Benjamin J. and {Petigura}, Erik A. and {Blunt}, Sarah and {Sinukoff}, Evan},
        title = "{RadVel: The Radial Velocity Modeling Toolkit}",
      journal = {\pasp},
         year = 2018,
        month = apr,
       volume = {130},
       number = {986},
        pages = {044504},
          doi = {10.1088/1538-3873/aaaaa8},
archivePrefix = {arXiv},
       eprint = {1801.01947},
 primaryClass = {astro-ph.IM},
       adsurl = {https://ui.adsabs.harvard.edu/abs/2018PASP..130d4504F}
}

@article{deka-szymankiewicz_penn_2018,
	title = {The {Penn} {State} - {Toruń} {Centre} for {Astronomy} {Planet} {Search} stars: {IV}. {Dwarfs} and the complete sample},
	volume = {615},
	copyright = {https://www.edpsciences.org/en/authors/copyright-and-licensing},
	issn = {0004-6361, 1432-0746},
	shorttitle = {The {Penn} {State} - {Toruń} {Centre} for {Astronomy} {Planet} {Search} stars},
	url = {https://www.aanda.org/10.1051/0004-6361/201731696},
	doi = {10.1051/0004-6361/201731696},
	urldate = {2025-10-17},
	journal = {Astronomy \& Astrophysics},
	author = {Deka-Szymankiewicz, B. and Niedzielski, A. and Adamczyk, M. and Adamów, M. and Nowak, G. and Wolszczan, A.},
	month = jul,
	year = {2018},
	pages = {A31},
}

@ARTICLE{2018AJ....156..117F,
       author = {{Fekel}, Francis C. and {Willmarth}, Daryl W. and {Abt}, Helmut A. and {Pourbaix}, Dimitri},
        title = "{Spectroscopic Orbits for Late-type Stars. II}",
      journal = {\aj},
         year = 2018,
        month = sep,
       volume = {156},
       number = {3},
          eid = {117},
        pages = {117},
          doi = {10.3847/1538-3881/aad3c1},
       adsurl = {https://ui.adsabs.harvard.edu/abs/2018AJ....156..117F}
}

@ARTICLE{2018MNRAS.478.4720G,
       author = {{G{\"u}nther}, Maximilian N. and {Queloz}, Didier and {Gillen}, Edward and {Delrez}, Laetitia and {Bouchy}, Fran{\c{c}}ois and {McCormac}, James and {Smalley}, Barry and {Almleaky}, Yaseen and {Armstrong}, David J. and {Bayliss}, Daniel and {Burdanov}, Artem and {Burleigh}, Matthew and {Cabrera}, Juan and {Casewell}, Sarah L. and {Cooke}, Benjamin F. and {Csizmadia}, Szil{\'a}rd and {Ducrot}, Elsa and {Eigm{\"u}ller}, Philipp and {Erikson}, Anders and {G{\"a}nsicke}, Boris T. and {Gibson}, Neale P. and {Gillon}, Micha{\"e}l and {Goad}, Michael R. and {Jehin}, Emmanu{\"e}l and {Jenkins}, James S. and {Louden}, Tom and {Moyano}, Maximiliano and {Murray}, Catriona and {Pollacco}, Don and {Poppenhaeger}, Katja and {Rauer}, Heike and {Raynard}, Liam and {Smith}, Alexis M.~S. and {Sohy}, Sandrine and {Thompson}, Samantha J. and {Udry}, St{\'e}phane and {Watson}, Christopher A. and {West}, Richard G. and {Wheatley}, Peter J.},
        title = "{Unmasking the hidden NGTS-3Ab: a hot Jupiter in an unresolved binary system}",
      journal = {\mnras},
         year = 2018,
        month = aug,
       volume = {478},
       number = {4},
        pages = {4720-4737},
          doi = {10.1093/mnras/sty1193},
archivePrefix = {arXiv},
       eprint = {1805.01378},
 primaryClass = {astro-ph.EP},
       adsurl = {https://ui.adsabs.harvard.edu/abs/2018MNRAS.478.4720G}
}

@ARTICLE{2019A&A...623A..72K,
       author = {{Kervella}, Pierre and {Arenou}, Fr{\'e}d{\'e}ric and {Mignard}, Fran{\c{c}}ois and {Th{\'e}venin}, Fr{\'e}d{\'e}ric},
        title = "{Stellar and substellar companions of nearby stars from Gaia DR2. Binarity from proper motion anomaly}",
      journal = {\aap},
         year = 2019,
        month = mar,
       volume = {623},
          eid = {A72},
        pages = {A72},
          doi = {10.1051/0004-6361/201834371},
archivePrefix = {arXiv},
       eprint = {1811.08902},
 primaryClass = {astro-ph.SR},
       adsurl = {https://ui.adsabs.harvard.edu/abs/2019A&A...623A..72K}
}

@ARTICLE{2019Natur.567..200P,
       author = {{Pietrzy{\'n}ski}, G. and {Graczyk}, D. and {Gallenne}, A. and {Gieren}, W. and {Thompson}, I.~B. and {Pilecki}, B. and {Karczmarek}, P. and {G{\'o}rski}, M. and {Suchomska}, K. and {Taormina}, M. and {Zgirski}, B. and {Wielg{\'o}rski}, P. and {Ko{\l}aczkowski}, Z. and {Konorski}, P. and {Villanova}, S. and {Nardetto}, N. and {Kervella}, P. and {Bresolin}, F. and {Kudritzki}, R.~P. and {Storm}, J. and {Smolec}, R. and {Narloch}, W.},
        title = "{A distance to the Large Magellanic Cloud that is precise to one per cent}",
      journal = {\nat},
         year = 2019,
        month = mar,
       volume = {567},
       number = {7747},
        pages = {200-203},
          doi = {10.1038/s41586-019-0999-4},
archivePrefix = {arXiv},
       eprint = {1903.08096},
 primaryClass = {astro-ph.GA},
       adsurl = {https://ui.adsabs.harvard.edu/abs/2019Natur.567..200P}
}

@ARTICLE{2019A&A...631A.125K,
       author = {{Kiefer}, F. and {H{\'e}brard}, G. and {Sahlmann}, J. and {Sousa}, S.~G. and {Forveille}, T. and {Santos}, N. and {Mayor}, M. and {Deleuil}, M. and {Wilson}, P.~A. and {Dalal}, S. and {D{\'\i}az}, R.~F. and {Henry}, G.~W. and {Hagelberg}, J. and {Hobson}, M.~J. and {Demangeon}, O. and {Bourrier}, V. and {Delfosse}, X. and {Arnold}, L. and {Astudillo-Defru}, N. and {Beuzit}, J.-L. and {Boisse}, I. and {Bonfils}, X. and {Borgniet}, S. and {Bouchy}, F. and {Courcol}, B. and {Ehrenreich}, D. and {Hara}, N. and {Lagrange}, A.-M. and {Lovis}, C. and {Montagnier}, G. and {Moutou}, C. and {Pepe}, F. and {Perrier}, C. and {Rey}, J. and {Santerne}, A. and {S{\'e}gransan}, D. and {Udry}, S. and {Vidal-Madjar}, A.},
        title = "{Detection and characterisation of 54 massive companions with the SOPHIE spectrograph. Seven new brown dwarfs and constraints on the brown dwarf desert}",
      journal = {\aap},
         year = 2019,
        month = nov,
       volume = {631},
          eid = {A125},
        pages = {A125},
          doi = {10.1051/0004-6361/201935113},
archivePrefix = {arXiv},
       eprint = {1909.00739},
 primaryClass = {astro-ph.SR},
       adsurl = {https://ui.adsabs.harvard.edu/abs/2019A&A...631A.125K}
}

@ARTICLE{2020SciPy-NMeth,
  author  = {Virtanen, Pauli and Gommers, Ralf and Oliphant, Travis E. and
            Haberland, Matt and Reddy, Tyler and Cournapeau, David and
            Burovski, Evgeni and Peterson, Pearu and Weckesser, Warren and
            Bright, Jonathan and {van der Walt}, St{\'e}fan J. and
            Brett, Matthew and Wilson, Joshua and Millman, K. Jarrod and
            Mayorov, Nikolay and Nelson, Andrew R. J. and Jones, Eric and
            Kern, Robert and Larson, Eric and Carey, C J and
            Polat, {\.I}lhan and Feng, Yu and Moore, Eric W. and
            {VanderPlas}, Jake and Laxalde, Denis and Perktold, Josef and
            Cimrman, Robert and Henriksen, Ian and Quintero, E. A. and
            Harris, Charles R. and Archibald, Anne M. and
            Ribeiro, Ant{\^o}nio H. and Pedregosa, Fabian and
            {van Mulbregt}, Paul and {SciPy 1.0 Contributors}},
  title   = {{{SciPy} 1.0: Fundamental Algorithms for Scientific
            Computing in Python}},
  journal = {Nature Methods},
  year    = {2020},
  volume  = {17},
  pages   = {261--272},
  adsurl  = {https://rdcu.be/b08Wh},
  doi     = {10.1038/s41592-019-0686-2},
}

@dataset{2022yCat.1357....0G,
       author = {{Gaia Collaboration}},
        title = "{VizieR Online Data Catalog: Gaia DR3 Part 3. Non-single stars (Gaia Collaboration, 2022)}",
 howpublished = {VizieR On-line Data Catalog: I/357.  Originally published in: 2023A\&A...674A..34G},
         year = 2022,
        month = may,
          eid = {I/357},
       adsurl = {https://ui.adsabs.harvard.edu/abs/2022yCat.1357....0G}
}

@ARTICLE{2022AJ....164..196W,
       author = {{Winn}, Joshua N.},
        title = "{Joint Constraints on Exoplanetary Orbits from Gaia DR3 and Doppler Data}",
      journal = {\aj},
         year = 2022,
        month = nov,
       volume = {164},
       number = {5},
          eid = {196},
        pages = {196},
          doi = {10.3847/1538-3881/ac9126},
archivePrefix = {arXiv},
       eprint = {2209.05516},
 primaryClass = {astro-ph.EP},
       adsurl = {https://ui.adsabs.harvard.edu/abs/2022AJ....164..196W}
}

@ARTICLE{2023ApJ...952..128C,
       author = {{Chae}, Kyu-Hyun},
        title = "{Breakdown of the Newton-Einstein Standard Gravity at Low Acceleration in Internal Dynamics of Wide Binary Stars}",
      journal = {\apj},
         year = 2023,
        month = aug,
       volume = {952},
       number = {2},
          eid = {128},
        pages = {128},
          doi = {10.3847/1538-4357/ace101},
archivePrefix = {arXiv},
       eprint = {2305.04613},
 primaryClass = {astro-ph.GA},
       adsurl = {https://ui.adsabs.harvard.edu/abs/2023ApJ...952..128C}
}

@ARTICLE{GaiaCollaboration2023NSS,
       author = {{Gaia Collaboration} and {Arenou}, F. and {Babusiaux}, C. and {Barstow}, M.~A. and {Faigler}, S. and {Jorissen}, A. and {Kervella}, P. and {Mazeh}, T. and {Mowlavi}, N. and {Panuzzo}, P. and {Sahlmann}, J. and {Shahaf}, S. and {Sozzetti}, A. and {Bauchet}, N. and {Damerdji}, Y. and {Gavras}, P. and {Giacobbe}, P. and {Gosset}, E. and {Halbwachs}, J.-L. and {Holl}, B. and {Lattanzi}, M.~G. and {Leclerc}, N. and {Morel}, T. and {Pourbaix}, D. and {Re Fiorentin}, P. and {Sadowski}, G. and {S{\'e}gransan}, D. and {Siopis}, C. and {Teyssier}, D. and {Zwitter}, T. and {Planquart}, L. and {Brown}, A.~G.~A. and {Vallenari}, A. and {Prusti}, T. and {de Bruijne}, J.~H.~J. and {Biermann}, M. and {Creevey}, O.~L. and {Ducourant}, C. and {Evans}, D.~W. and {Eyer}, L. and {Guerra}, R. and {Hutton}, A. and {Jordi}, C. and {Klioner}, S.~A. and {Lammers}, U.~L. and {Lindegren}, L. and {Luri}, X. and {Mignard}, F. and {Panem}, C. and {Randich}, S. and {Sartoretti}, P. and {Soubiran}, C. and {Tanga}, P. and {Walton}, N.~A. and {Bailer-Jones}, C.~A.~L. and {Bastian}, U. and {Drimmel}, R. and {Jansen}, F. and {Katz}, D. and {van Leeuwen}, F. and {Bakker}, J. and {Cacciari}, C. and {Casta{\~n}eda}, J. and {De Angeli}, F. and {Fabricius}, C. and {Fouesneau}, M. and {Fr{\'e}mat}, Y. and {Galluccio}, L. and {Guerrier}, A. and {Heiter}, U. and {Masana}, E. and {Messineo}, R. and {Nicolas}, C. and {Nienartowicz}, K. and {Pailler}, F. and {Riclet}, F. and {Roux}, W. and {Seabroke}, G.~M. and {Sordo}, R. and {Th{\'e}venin}, F. and {Gracia-Abril}, G. and {Portell}, J. and {Altmann}, M. and {Andrae}, R. and {Audard}, M. and {Bellas-Velidis}, I. and {Benson}, K. and {Berthier}, J. and {Blomme}, R. and {Burgess}, P.~W. and {Busonero}, D. and {Busso}, G. and {C{\'a}novas}, H. and {Carry}, B. and {Cellino}, A. and {Cheek}, N. and {Clementini}, G. and {Davidson}, M. and {de Teodoro}, P. and {Nu{\~n}ez Campos}, M. and {Delchambre}, L. and {Dell'Oro}, A. and {Esquej}, P. and {Fern{\'a}ndez-Hern{\'a}ndez}, J. and {Fraile}, E. and {Garabato}, D. and {Garc{\'\i}a-Lario}, P. and {Haigron}, R. and {Hambly}, N.~C. and {Harrison}, D.~L. and {Hern{\'a}ndez}, J. and {Hestroffer}, D. and {Hodgkin}, S.~T. and {Jan{\ss}en}, K. and {Jevardat de Fombelle}, G. and {Jordan}, S. and {Krone-Martins}, A. and {Lanzafame}, A.~C. and {L{\"o}ffler}, W. and {Marchal}, O. and {Marrese}, P.~M. and {Moitinho}, A. and {Muinonen}, K. and {Osborne}, P. and {Pancino}, E. and {Pauwels}, T. and {Recio-Blanco}, A. and {Reyl{\'e}}, C. and {Riello}, M. and {Rimoldini}, L. and {Roegiers}, T. and {Rybizki}, J. and {Sarro}, L.~M. and {Smith}, M. and {Utrilla}, E. and {van Leeuwen}, M. and {Abbas}, U. and {{\'A}brah{\'a}m}, P. and {Abreu Aramburu}, A. and {Aerts}, C. and {Aguado}, J.~J. and {Ajaj}, M. and {Aldea-Montero}, F. and {Altavilla}, G. and {{\'A}lvarez}, M.~A. and {Alves}, J. and {Anders}, F. and {Anderson}, R.~I. and {Anglada Varela}, E. and {Antoja}, T. and {Baines}, D. and {Baker}, S.~G. and {Balaguer-N{\'u}{\~n}ez}, L. and {Balbinot}, E. and {Balog}, Z. and {Barache}, C. and {Barbato}, D. and {Barros}, M. and {Bartolom{\'e}}, S. and {Bassilana}, J.-L. and {Becciani}, U. and {Bellazzini}, M. and {Berihuete}, A. and {Bernet}, M. and {Bertone}, S. and {Bianchi}, L. and {Binnenfeld}, A. and {Blanco-Cuaresma}, S. and {Blazere}, A. and {Boch}, T. and {Bombrun}, A. and {Bossini}, D. and {Bouquillon}, S. and {Bragaglia}, A. and {Bramante}, L. and {Breedt}, E. and {Bressan}, A. and {Brouillet}, N. and {Brugaletta}, E. and {Bucciarelli}, B. and {Burlacu}, A. and {Butkevich}, A.~G. and {Buzzi}, R. and {Caffau}, E. and {Cancelliere}, R. and {Cantat-Gaudin}, T. and {Carballo}, R. and {Carlucci}, T. and {Carnerero}, M.~I. and {Carrasco}, J.~M. and {Casamiquela}, L. and {Castellani}, M. and {Castro-Ginard}, A. and {Chaoul}, L. and {Charlot}, P. and {Chemin}, L. and {Chiaramida}, V. and {Chiavassa}, A. and {Chornay}, N. and {Comoretto}, G.},
        title = "{Gaia Data Release 3. Stellar multiplicity, a teaser for the hidden treasure}",
      journal = {\aap},
         year = 2023,
        month = jun,
       volume = {674},
          eid = {A34},
        pages = {A34},
          doi = {10.1051/0004-6361/202243782},
archivePrefix = {arXiv},
       eprint = {2206.05595},
 primaryClass = {astro-ph.SR},
       adsurl = {https://ui.adsabs.harvard.edu/abs/2023A&A...674A..34G}
}

@ARTICLE{GaiaCollaboration2023DR3Summary,
       author = {{Gaia Collaboration} and {Vallenari}, A. and {Brown}, A.~G.~A. and {Prusti}, T. and {de Bruijne}, J.~H.~J. and {Arenou}, F. and {Babusiaux}, C. and {Biermann}, M. and {Creevey}, O.~L. and {Ducourant}, C. and {Evans}, D.~W. and {Eyer}, L. and {Guerra}, R. and {Hutton}, A. and {Jordi}, C. and {Klioner}, S.~A. and {Lammers}, U.~L. and {Lindegren}, L. and {Luri}, X. and {Mignard}, F. and {Panem}, C. and {Pourbaix}, D. and {Randich}, S. and {Sartoretti}, P. and {Soubiran}, C. and {Tanga}, P. and {Walton}, N.~A. and {Bailer-Jones}, C.~A.~L. and {Bastian}, U. and {Drimmel}, R. and {Jansen}, F. and {Katz}, D. and {Lattanzi}, M.~G. and {van Leeuwen}, F. and {Bakker}, J. and {Cacciari}, C. and {Casta{\~n}eda}, J. and {De Angeli}, F. and {Fabricius}, C. and {Fouesneau}, M. and {Fr{\'e}mat}, Y. and {Galluccio}, L. and {Guerrier}, A. and {Heiter}, U. and {Masana}, E. and {Messineo}, R. and {Mowlavi}, N. and {Nicolas}, C. and {Nienartowicz}, K. and {Pailler}, F. and {Panuzzo}, P. and {Riclet}, F. and {Roux}, W. and {Seabroke}, G.~M. and {Sordo}, R. and {Th{\'e}venin}, F. and {Gracia-Abril}, G. and {Portell}, J. and {Teyssier}, D. and {Altmann}, M. and {Andrae}, R. and {Audard}, M. and {Bellas-Velidis}, I. and {Benson}, K. and {Berthier}, J. and {Blomme}, R. and {Burgess}, P.~W. and {Busonero}, D. and {Busso}, G. and {C{\'a}novas}, H. and {Carry}, B. and {Cellino}, A. and {Cheek}, N. and {Clementini}, G. and {Damerdji}, Y. and {Davidson}, M. and {de Teodoro}, P. and {Nu{\~n}ez Campos}, M. and {Delchambre}, L. and {Dell'Oro}, A. and {Esquej}, P. and {Fern{\'a}ndez-Hern{\'a}ndez}, J. and {Fraile}, E. and {Garabato}, D. and {Garc{\'\i}a-Lario}, P. and {Gosset}, E. and {Haigron}, R. and {Halbwachs}, J.-L. and {Hambly}, N.~C. and {Harrison}, D.~L. and {Hern{\'a}ndez}, J. and {Hestroffer}, D. and {Hodgkin}, S.~T. and {Holl}, B. and {Jan{\ss}en}, K. and {Jevardat de Fombelle}, G. and {Jordan}, S. and {Krone-Martins}, A. and {Lanzafame}, A.~C. and {L{\"o}ffler}, W. and {Marchal}, O. and {Marrese}, P.~M. and {Moitinho}, A. and {Muinonen}, K. and {Osborne}, P. and {Pancino}, E. and {Pauwels}, T. and {Recio-Blanco}, A. and {Reyl{\'e}}, C. and {Riello}, M. and {Rimoldini}, L. and {Roegiers}, T. and {Rybizki}, J. and {Sarro}, L.~M. and {Siopis}, C. and {Smith}, M. and {Sozzetti}, A. and {Utrilla}, E. and {van Leeuwen}, M. and {Abbas}, U. and {{\'A}brah{\'a}m}, P. and {Abreu Aramburu}, A. and {Aerts}, C. and {Aguado}, J.~J. and {Ajaj}, M. and {Aldea-Montero}, F. and {Altavilla}, G. and {{\'A}lvarez}, M.~A. and {Alves}, J. and {Anders}, F. and {Anderson}, R.~I. and {Anglada Varela}, E. and {Antoja}, T. and {Baines}, D. and {Baker}, S.~G. and {Balaguer-N{\'u}{\~n}ez}, L. and {Balbinot}, E. and {Balog}, Z. and {Barache}, C. and {Barbato}, D. and {Barros}, M. and {Barstow}, M.~A. and {Bartolom{\'e}}, S. and {Bassilana}, J.-L. and {Bauchet}, N. and {Becciani}, U. and {Bellazzini}, M. and {Berihuete}, A. and {Bernet}, M. and {Bertone}, S. and {Bianchi}, L. and {Binnenfeld}, A. and {Blanco-Cuaresma}, S. and {Blazere}, A. and {Boch}, T. and {Bombrun}, A. and {Bossini}, D. and {Bouquillon}, S. and {Bragaglia}, A. and {Bramante}, L. and {Breedt}, E. and {Bressan}, A. and {Brouillet}, N. and {Brugaletta}, E. and {Bucciarelli}, B. and {Burlacu}, A. and {Butkevich}, A.~G. and {Buzzi}, R. and {Caffau}, E. and {Cancelliere}, R. and {Cantat-Gaudin}, T. and {Carballo}, R. and {Carlucci}, T. and {Carnerero}, M.~I. and {Carrasco}, J.~M. and {Casamiquela}, L. and {Castellani}, M. and {Castro-Ginard}, A. and {Chaoul}, L. and {Charlot}, P. and {Chemin}, L. and {Chiaramida}, V. and {Chiavassa}, A. and {Chornay}, N. and {Comoretto}, G. and {Contursi}, G. and {Cooper}, W.~J. and {Cornez}, T. and {Cowell}, S. and {Crifo}, F. and {Cropper}, M. and {Crosta}, M. and {Crowley}, C. and {Dafonte}, C. and {Dapergolas}, A. and {David}, M. and {David}, P. and {de Laverny}, P. and {De Luise}, F. and {De March}, R.},
        title = "{Gaia Data Release 3. Summary of the content and survey properties}",
      journal = {\aap},
         year = 2023,
        month = jun,
       volume = {674},
          eid = {A1},
        pages = {A1},
          doi = {10.1051/0004-6361/202243940},
archivePrefix = {arXiv},
       eprint = {2208.00211},
 primaryClass = {astro-ph.GA},
       adsurl = {https://ui.adsabs.harvard.edu/abs/2023A&A...674A...1G}
}

@ARTICLE{2024CoSka..54b..43P,
       author = {{Pribulla}, T. and {Va{\v{n}}ko}, M. and {Kom{\v{z}}{\'\i}k}, R. and {Sivani{\v{c}}}, P.},
        title = "{High-resolution {\'e}chelle spectrograph at Skalnat{\'e} Pleso Observatory}",
      journal = {Contributions of the Astronomical Observatory Skalnate Pleso},
         year = 2024,
        month = feb,
       volume = {54},
       number = {2},
        pages = {43-46},
          doi = {10.31577/caosp.2024.54.2.43},
       adsurl = {https://ui.adsabs.harvard.edu/abs/2024CoSka..54b..43P}
}

@ARTICLE{2025A&A...700A.106T,
       author = {{Thebault}, P. and {Bonanni}, D.},
        title = "{A complete census of planet-hosting binaries}",
      journal = {\aap},
         year = 2025,
        month = aug,
       volume = {700},
          eid = {A106},
        pages = {A106},
          doi = {10.1051/0004-6361/202555457},
archivePrefix = {arXiv},
       eprint = {2506.18759},
 primaryClass = {astro-ph.EP},
       adsurl = {https://ui.adsabs.harvard.edu/abs/2025A&A...700A.106T}
}

@ARTICLE{2025A&A...695A..62S,
       author = {{Stefanov}, A.~K. and {Su{\'a}rez Mascare{\~n}o}, A. and {Gonz{\'a}lez Hern{\'a}ndez}, J.~I. and {Nari}, N. and {Rebolo}, R. and {Affer}, L. and {Micela}, G. and {Ribas}, I. and {Sozzetti}, A. and {Perger}, M. and {Pinamonti}, M. and {Damasso}, M. and {Maldonado}, J. and {Gonz{\'a}lez {\'A}lvarez}, E. and {Scandariato}, G.},
        title = "{HADES RV Programme with HARPS-N at TNG: XVI. A super-Earth in the habitable zone of the GJ 3998 multi-planet system}",
      journal = {\aap},
         year = 2025,
        month = mar,
       volume = {695},
          eid = {A62},
        pages = {A62},
          doi = {10.1051/0004-6361/202452630},
archivePrefix = {arXiv},
       eprint = {2503.08405},
 primaryClass = {astro-ph.EP},
       adsurl = {https://ui.adsabs.harvard.edu/abs/2025A&A...695A..62S}
}

@ARTICLE{2025arXiv250420825G,
       author = {{Giangrandi}, Edoardo and {R{\"u}ter}, Hannes R. and {Kunert}, Nina and {Emma}, Mattia and {Abac}, Adrian and {Adhikari}, Ananya and {Dietrich}, Tim and {Sagun}, Violetta and {Tichy}, Wolfgang and {Provid{\^e}ncia}, Constan{\c{c}}a},
        title = "{Numerical Relativity Simulations of Dark Matter Admixed Binary Neutron Stars}",
      journal = {arXiv e-prints},
         year = 2025,
        month = apr,
          eid = {arXiv:2504.20825},
        pages = {arXiv:2504.20825},
          doi = {10.48550/arXiv.2504.20825},
archivePrefix = {arXiv},
       eprint = {2504.20825},
 primaryClass = {astro-ph.HE},
       adsurl = {https://ui.adsabs.harvard.edu/abs/2025arXiv250420825G}
}

@ARTICLE{2025CoSka..55c..21B,
       author = {{Boffin}, H.~M.~J. and {Jones}, D.},
        title = "{The importance of binary stars}",
      journal = {Contributions of the Astronomical Observatory Skalnate Pleso},
         year = 2025,
        month = apr,
       volume = {55},
       number = {3},
        pages = {21-36},
          doi = {10.31577/caosp.2025.55.3.21},
archivePrefix = {arXiv},
       eprint = {2411.18470},
 primaryClass = {astro-ph.SR},
       adsurl = {https://ui.adsabs.harvard.edu/abs/2025CoSka..55c..21B}
}

@ARTICLE{2016A&A...585A..73N,
       author = {{Niedzielski}, A. and {Deka-Szymankiewicz}, B. and {Adamczyk}, M. and {Adam{\'o}w}, M. and {Nowak}, G. and {Wolszczan}, A.},
        title = "{The Penn State - Toru{\'n} Centre for Astronomy Planet Search stars}",
      journal = {\aap},
         year = 2016,
        month = jan,
       volume = {585},
          eid = {A73},
        pages = {A73},
          doi = {10.1051/0004-6361/201527362},
       adsurl = {https://ui.adsabs.harvard.edu/abs/2016A&A...585A..73N}
}

@ARTICLE{2000PASP..112..137M,
       author = {{Marcy}, Geoffrey W. and {Butler}, R. Paul},
        title = "{Planets Orbiting Other Suns}",
      journal = {\pasp},
         year = 2000,
        month = feb,
       volume = {112},
       number = {768},
        pages = {137-140},
          doi = {10.1086/316516},
       adsurl = {https://ui.adsabs.harvard.edu/abs/2000PASP..112..137M}
}

@ARTICLE{2025AcA....75...63N,
       author = {{Niedzielski}, A. and {Jaros}, R. and {Srivastava}, D. and {Adam{\'o}w}, M. and {Wolszczan}, A. and {Villaver}, E. and {Maciejewski}, G. and {Deka-Szymankiewicz}, B.},
        title = "{Low-Mass Companions to Nine Stars}",
      journal = {\actaa},
         year = 2025,
        month = nov,
       volume = {75},
       number = {2},
        pages = {63-95},
          doi = {10.32023/0001-5237/75.2.1},
archivePrefix = {arXiv},
       eprint = {2509.22127},
 primaryClass = {astro-ph.SR},
       adsurl = {https://ui.adsabs.harvard.edu/abs/2025AcA....75...63N}
}

\begin{appendix}
\onecolumn

\begin{table*}[ht!]
\section{Additional material}
    \centering
    \caption{Presentation of RV information from the HET, CAFE, SPO, and TNG instruments. }
    \label{HET_obs}
    \begin{tabular}{l|rrrrrrr}
         TYC & First & Last & Span [d] & Amplitude [m s$^{-1}$] & RV error median [m s$^{-1}$] & N & Instrument\\
         \hline
0650-01471 & 2005-12-27 & 2012-12-07 & 2537.06 & 8772.43 & 9.31 & 27 & HET \\
0650-01471 & 2024-10-01 & 2024-12-07 & 67.73 & 269.62 & 566.27 & 5 & SPO \\
0650-01471 & 2006-11-06 & 2009-01-25 & 810.85 & 8618.86 & 1.05 & 9 & TNG \\
0697-01743 & 2005-12-31 & 2013-01-25 & 2581.93 & 8803.44 & 6.69 & 33 & HET \\
0697-01743 & 2024-11-08 & 2024-12-14 & 35.99 & 210.43 & 219.18 & 8 & SPO \\
0749-00973 & 2006-01-10 & 2013-02-03 & 2580.94 & 5744.55 & 9.52 & 40 & HET \\
0749-00973 & 2017-10-19 & 2018-01-23 & 95.90 & 225.22 & 2.06 & 4 & TNG \\
0749-00973 & 2024-11-08 & 2024-12-14 & 35.95 & 258.92 & 1017.46 & 8 & SPO \\
0870-00084 & 2004-03-07 & 2013-05-22 & 3362.82 & 3988.32 & 19.66 & 16 & HET \\
0870-00084 & 2025-03-05 & 2025-03-27 & 22.04 & 219.85 & 368.06 & 5 & SPO \\
0870-00937 & 2004-01-22 & 2011-03-28 & 2621.83 & 2149.02 & 6.02 & 29 & HET \\
0870-00937 & 2025-02-04 & 2025-04-12 & 67.87 & 158.83 & 261.82 & 6 & SPO \\
1931-01040 & 2005-11-27 & 2013-05-06 & 2716.79 & 7965.03 & 16.12 & 14 & HET \\
1931-01040 & 2013-01-09 & 2015-01-06 & 727.06 & 7837.10 & 126.82 & 12 & CAFE \\
1931-01040 & 2025-03-05 & 2025-03-27 & 21.97 & 807.10 & 885.37 & 4 & SPO \\
2267-00101 & 2006-06-12 & 2013-06-28 & 2572.96 & 6951.04 & 9.04 & 17 & HET \\
2267-00101 & 2012-11-30 & 2015-02-12 & 803.92 & 6305.68 & 1.10 & 14 & TNG \\
2267-00101 & 2013-01-08 & 2015-11-16 & 1041.94 & 6462.00 & 27.00 & 17 & CAFE \\
2822-01643 & 2006-01-11 & 2013-02-14 & 2590.92 & 4361.68 & 5.27 & 15 & HET \\
2822-01643 & 2013-01-07 & 2015-01-29 & 751.91 & 1520.24 & 110.72 & 12 & CAFE \\
3018-01050 & 2004-01-30 & 2013-04-27 & 3374.74 & 9190.12 & 6.80 & 19 & HET \\
3018-01050 & 2013-01-06 & 2015-01-28 & 752.09 & 3187.37 & 86.04 & 13 & CAFE \\
3314-01371 & 2004-10-30 & 2013-01-14 & 2997.81 & 7363.13 & 7.95 & 15 & HET \\
3314-01371 & 2024-10-16 & 2024-12-13 & 58.89 & 216.51 & 496.72 & 7 & SPO \\
3314-01371 & 2013-01-28 & 2018-01-23 & 1820.05 & 3407.34 & 2.96 & 6 & TNG \\
3318-00789 & 2004-02-01 & 2012-12-19 & 3243.89 & 9457.44 & 10.50 & 16 & HET \\
3318-00789 & 2024-10-16 & 2024-12-13 & 58.89 & 2761.04 & 465.92 & 6 & SPO \\
3318-01427 & 2006-10-21 & 2012-02-02 & 1929.97 & 4888.19 & 6.46 & 34 & HET \\
3318-01427 & 2024-10-18 & 2024-12-12 & 55.00 & 376.85 & 445.59 & 5 & SPO \\
3318-01538 & 2005-01-03 & 2012-12-30 & 2917.99 & 9022.10 & 5.97 & 18 & HET \\
3318-01538 & 2013-01-06 & 2015-01-29 & 753.87 & 10666.83 & 139.79 & 12 & CAFE \\
3318-01538 & 2024-10-16 & 2024-12-08 & 53.61 & 2617.46 & 297.14 & 5 & SPO \\
3319-00172 & 2004-02-06 & 2010-11-15 & 2474.22 & 12398.08 & 13.12 & 13 & HET \\
3319-00172 & 2024-10-01 & 2024-11-26 & 55.97 & 472.82 & 317.32 & 4 & SPO \\
3451-01449 & 2005-12-19 & 2013-06-26 & 2745.72 & 12392.69 & 6.51 & 9 & HET \\
3451-01449 & 2013-01-06 & 2015-01-29 & 752.89 & 10983.60 & 66.41 & 12 & CAFE \\
3667-01636 & 2005-09-02 & 2012-12-28 & 2673.71 & 5857.93 & 6.33 & 12 & HET \\
3667-01636 & 2013-01-09 & 2015-01-28 & 748.96 & 4511.08 & 86.29 & 10 & CAFE \\
3667-01636 & 2025-11-18 & 2025-11-30 & 12.12 & 6.91 & 255.45 & 2 & SPO \\
    \end{tabular} 
    \tablefoot{The "First" and "Last" columns show measurements in MJD time. Span column is the time between the first and last measurement. Amplitude is the difference between the minimum and maximum RV values, with the median value of uncertainty in the next column. The last column tells the number of RV epochs used for data analysis. }
\end{table*}

\begin{figure*}
    \centering
    \includegraphics[width=0.95\textwidth]{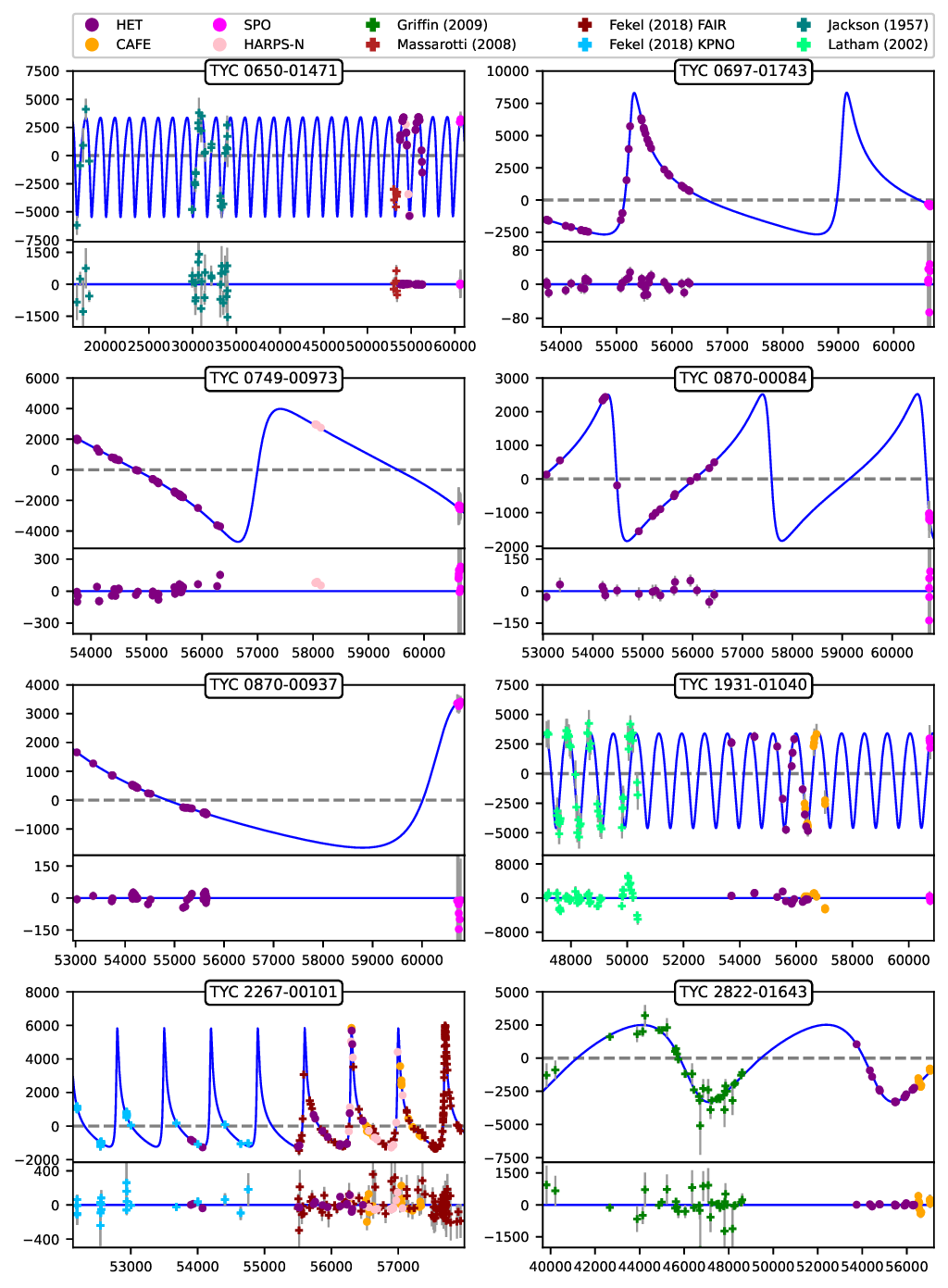}
        \caption{Part 1 of the RV time series. The dots on the figures represent our  data, where crosses represent data collected from literature. Blue lines represent fitted model. The y-axis is in units of m s$^{-1}$, and the x-axis is time in modified Julian date (MJD = JD - 2400000.5).} 
        \label{RV_models}
\end{figure*}
\begin{figure*}
    \centering
    \ContinuedFloat %
    \includegraphics[width=0.95\textwidth]{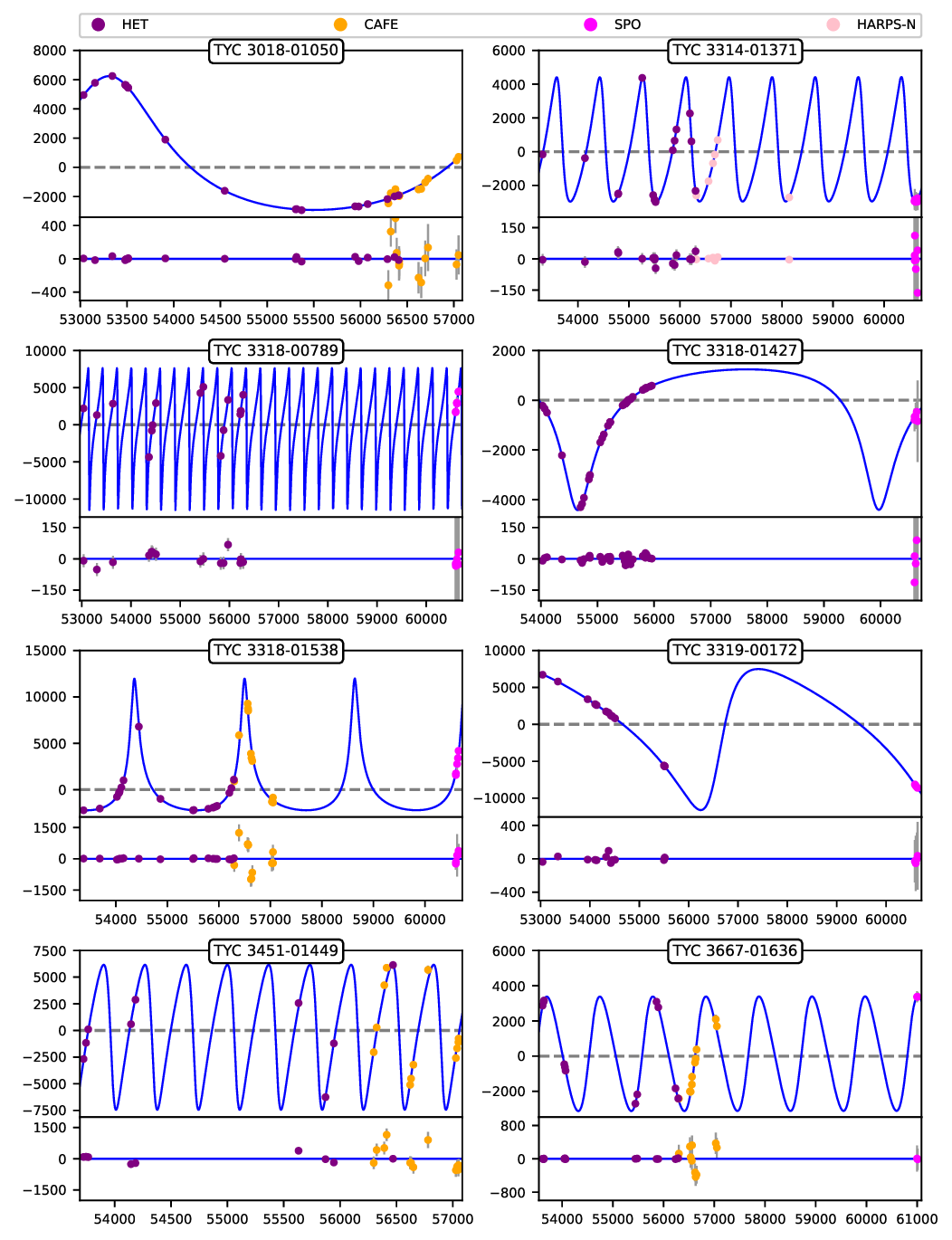}
    \caption{Part 2 of the RV time series.}
\end{figure*}

\begin{figure}
    \centering
    \includegraphics[width=0.8\textwidth]{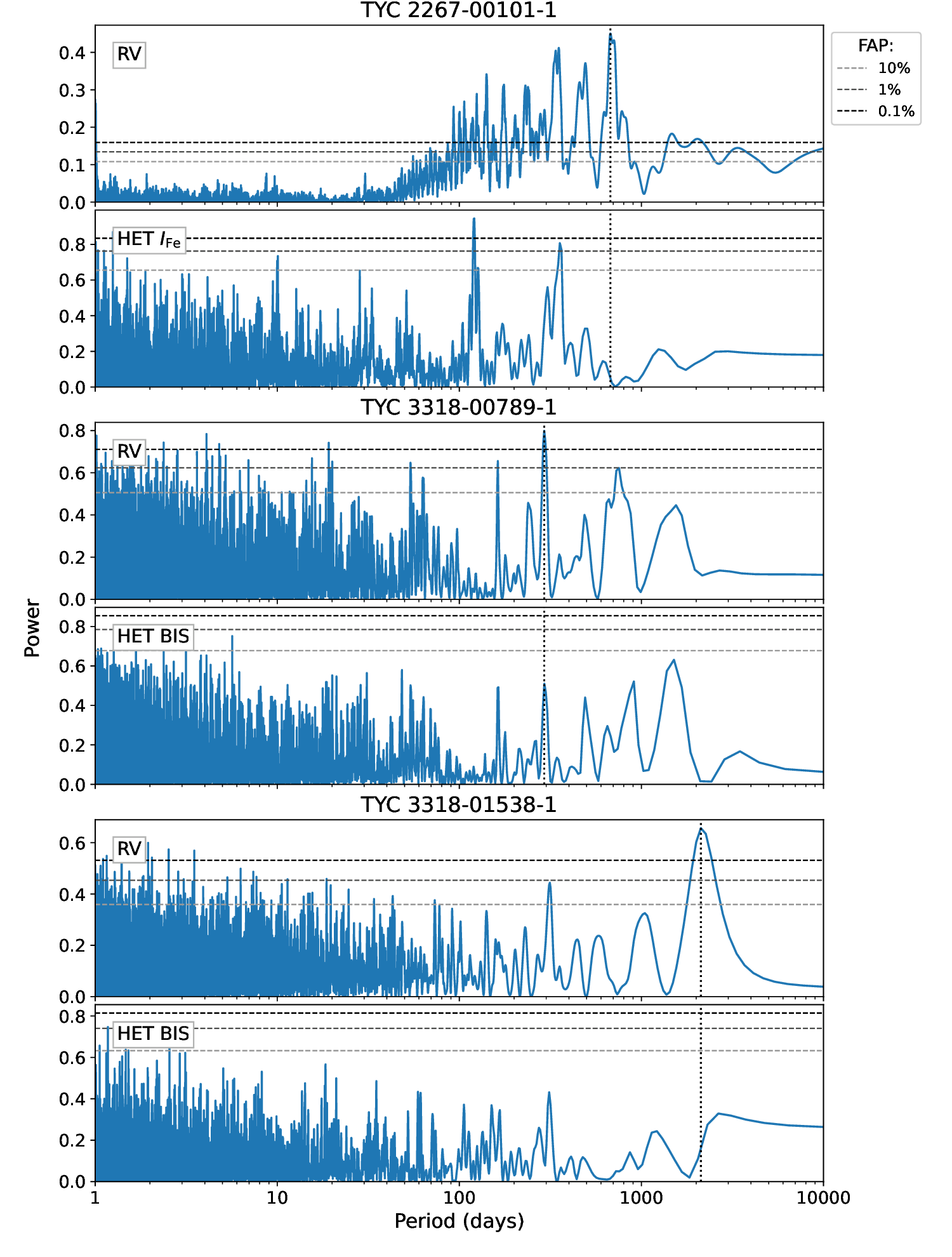}
    \caption{GLS periodograms of RV and the correlated HET activity indicators: $I_{\mathrm{Fe}}$ for TYC~2267-00101-1, and BIS for TYC~3318-00789-1 and TYC~3318-01538-1. Dotted vertical lines correspond to the strongest RV period. Activity panels of TYC~2267-00101-1 and TYC~3318-01538-1 do not show their peak around the dominant RV period. The BIS periodogram of TYC~3318-00789-1 shows weak peak which is below the FAP threshold, near the RV period.}
    \label{rv_activity}
\end{figure}

\begin{figure}
    \centering
    \includegraphics[width=1\linewidth]{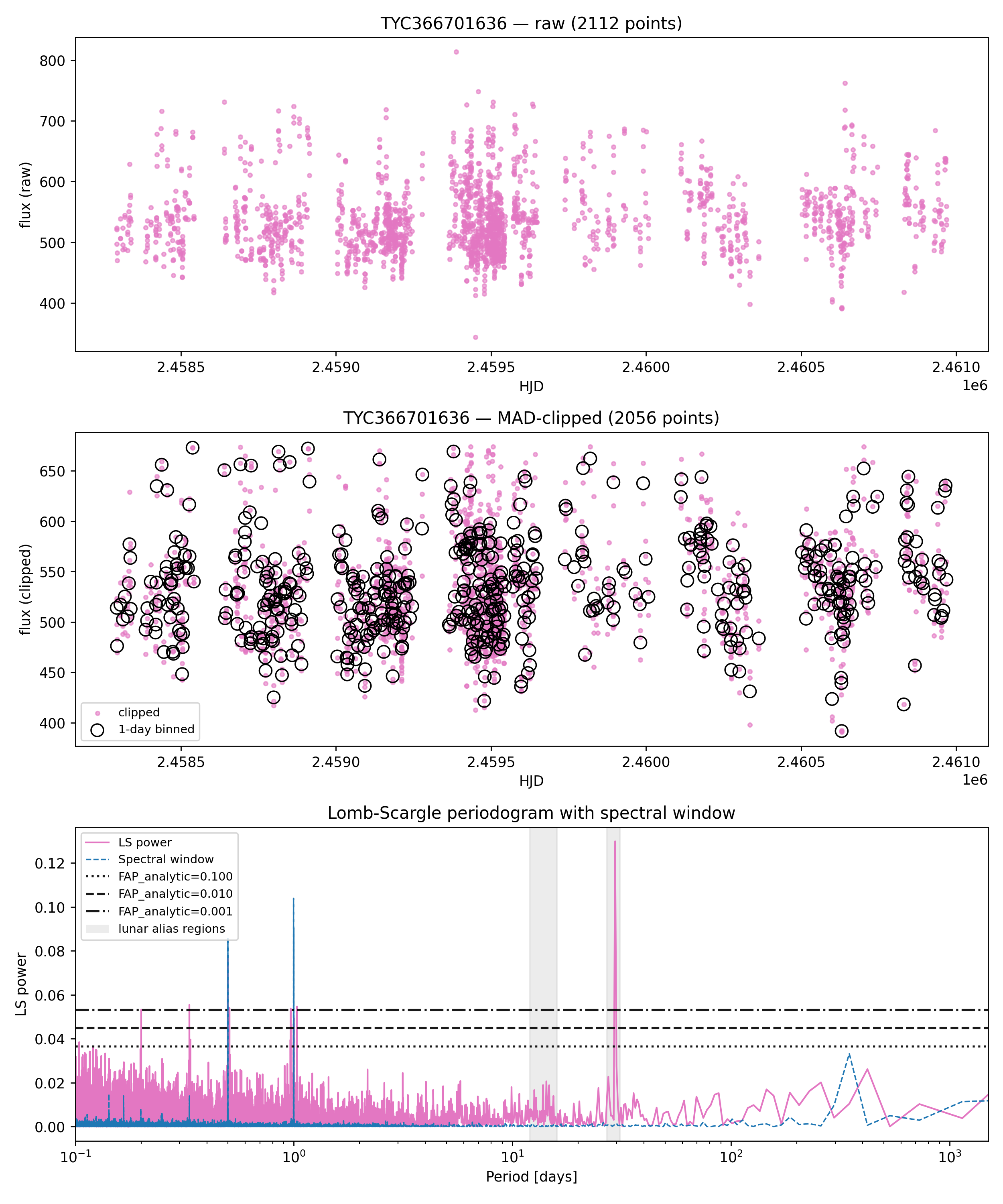}
    \caption{GLS periodogram for one of our targets TYC 3667-01636 in the ASAS-SN g band. The periodogram shows no significant peaks after taking into account aforementioned aliases, power peaks from trends and observational gaps. The lunar aliases appear in nearly every dataset and further strengthen the idea that these periods should be treated with caution in ASAS-SN datasets.}
    \label{photometry}
\end{figure}

\end{appendix}

\end{document}